\documentclass[aps,prd,preprint,groupedaddress]{revtex4-2}
\usepackage[utf8]{inputenc}
\usepackage[T1]{fontenc}
\usepackage{graphicx}
\usepackage{enumitem}
\usepackage{natbib}
\usepackage{aps_macros}
\usepackage{gensymb}
\usepackage{amsmath}
\usepackage{amssymb}
\usepackage{mathtools}
\usepackage{slashed}
\usepackage{float}

\begin{document}

\title{Indirect dark matter searches at ultrahigh energy neutrino detectors}

\author{Claire Gu\'epin}
\email{cguepin@umd.edu}
\affiliation{Department of Astronomy, University of Maryland, College Park, MD, USA}

\author{Roberto Aloisio}
\affiliation{Gran Sasso Science Institute, L'Aquila, Italy}

\author{Luis A. Anchordoqui}
\affiliation{Department of Physics and Astronomy, Lehman College, City University of New York,  NY NY 10468, USA}
\affiliation{Department of Astrophysics, American Museum of Natural History, NY 10024, USA}

\author{Austin Cummings}
\affiliation{Gran Sasso Science Institute, L'Aquila, Italy}

\author{John F. Krizmanic}
\affiliation{NASA Goddard Space Flight Center, Greenbelt, MD, USA}
\affiliation{Center for Space Science \& Technology, University of Maryland, Baltimore County, Baltimore, MD, USA}

\author{Angela V. Olinto}
\affiliation{Department of Astronomy \& Astrophysics, KICP, EFI, The University of Chicago, Chicago, IL 60637, USA}

\author{Mary Hall Reno}
\affiliation{Department of Physics and Astronomy, University of Iowa, Iowa City, IA, USA}

\author{Tonia M. Venters}
\affiliation{{NASA Goddard Space Flight Center, Greenbelt, MD, USA}}

\date{\today}

\begin{abstract}
    \noindent High to ultrahigh energy neutrino detectors can uniquely probe the properties of dark matter $\chi$ by searching for the secondary products produced through annihilation and/or decay processes. We evaluate the sensitivities to dark matter thermally averaged annihilation cross section $\langle\sigma v\rangle$ and partial decay width into neutrinos $\Gamma_{\chi\rightarrow\nu\bar{\nu}}$ (in the mass scale $10^7 \leq m_\chi/{\rm GeV} \leq  10^{15}$) for next generation observatories like POEMMA and GRAND. We show that in the range $ 10^7 \leq m_\chi/{\rm GeV} \leq 10^{11}$, space-based Cherenkov detectors like POEMMA have the advantage of full-sky coverage and rapid slewing, enabling an optimized dark matter observation strategy focusing on the Galactic center. We also show that ground-based radio detectors such as GRAND can achieve high sensitivities and high duty cycles in radio quiet areas. We compare the sensitivities of next generation neutrino experiments with existing constraints from IceCube and updated 90\% C.L. upper limits on  $\langle\sigma v\rangle$ and $\Gamma_{\chi\rightarrow\nu\bar{\nu}}$ using results from the Pierre Auger Collaboration and ANITA. We show that in the range $ 10^7 \leq m_\chi/{\rm GeV} \leq 10^{11}$ POEMMA and GRAND10k will improve the neutrino sensitivity to particle dark matter by factors of 2 to 10 over existing limits, whereas GRAND200k will improve this sensitivity by two orders of magnitude. In the range $10^{11} \leq m_\chi/{\rm GeV} \leq 10^{15}$, POEMMA's fluorescence observation mode will achieve an unprecedented sensitivity to dark matter properties. Finally, we highlight the importance of the uncertainties related to the dark matter distribution in the Galactic halo, using the latest fit and estimates of the Galactic parameters.
\end{abstract}

\maketitle

\section{Introduction}

The evidence of dark matter is compelling at various astrophysical scales, from Galactic scales to cosmological scales \citep[e.g.][for reviews]{Jungman:1995df, BHS05, Feng:2010gw, Garrett11, Bertone:2016nfn}. Following the first discoveries of stellar velocity anomalies in our Galaxy \citep{Oort32} and galaxy velocity dispersion anomalies in galaxy clusters \citep{Zwicky33,Zwicky37}, the existence of a dark matter component was firmly established by a variety of probes, such as the extensive study of galaxy rotation curves \citep{Rubin85,Begeman91}, gravitational lensing observations of galaxy clusters \citep[e.g.][]{Bergmann90}, weak gravitational lensing and X-ray observations of collisions between galaxy clusters \citep[e.g.][]{Clowe06,Bradac08}, observations of dwarf galaxies in galaxy clusters \citep[e.g.][]{Mateo08,Penny09}, observations of the cosmic microwave background temperature fluctuations \citep{Smoot92,Jarosik11}, observations of large-scale structures \citep{Percival10} and simulations of large scale structure formation \citep{DiMatteo08,BoylanKolchin09}.

Despite the extensive evidence for the existence of non-baryonic dark matter (DM), representing nearly $84\%$ of the matter density in the Universe, its nature is still elusive. A large number of candidates have been proposed, such as sterile neutrinos, axions, supersymmetric candidates such as neutralinos, sneutrinos, gravitinos and axinos, light scalar dark matter, dark matter from Little Higgs models, Kaluza-Klein states, superheavy dark matter, and many more~\citep{Ellis00,Feng:2010gw}. The diversity of possible particle candidates requires a balanced program based on four-pillar strategies for dark matter detection~\citep{Feng:2010gw, Bergstrom00, Klasen15, MarrodanUndagoitia16, Penning18}:
\begin{itemize}[noitemsep,topsep=0pt]
\item Collider experiments that elucidate the particle properties of DM. DM could be produced in the scattering of standard model (SM) particles. Although the DM particles would be undetectable they are typically accompanied by related production mechanisms, e.g., \mbox{${\rm SM}~{\rm SM} \to {\rm DM}~{\rm DM}  + \{\rm SM\}$}, where $\{{\rm SM}\}$ denotes one or more SM particles. 
\item Direct detection experiments that look for DM interacting in the lab. DM can scatter off SM particles via ${\rm DM}~{\rm SM} \to {\rm DM}~{\rm SM}$ interactions, depositing energy that could be detected by sensitive, low background experiments.
\item Indirect detection experiments that connect lab signals to DM in the galactic halos. DM can annihilate  ${\rm DM}~{\rm DM} \to {\rm SM}~{\rm SM}$ or decay ${\rm DM} \to {\rm SM}~{\rm SM}$, and the annihilation/decay products could be detected.
\item Astrophysical probes that determine how DM scattering ${\rm DM}~{\rm DM} \to {\rm DM}~{\rm DM}$ has shaped the evolution of large-scale structures in the Universe.
\end{itemize}
In this paper we focus attention on indirect detection of dark matter particles by searching for high- and ultrahigh-energy neutrinos. Before proceeding, we pause to describe two caveats of our analysis.

It is well-known that the SM of electroweak interactions includes three left-handed neutrino fields $\nu_{\alpha L}$, which accompany the three families of charged leptons $\ell_{\alpha L}$ in the $SU(2)_L$ lepton doublet $L_\alpha= (\nu_{\alpha L}, \ell_{\alpha L})^T$, where $\alpha = e,\mu,\tau$. Because SM neutrinos only interact through weak interactions the right-handed fields $\nu_{\alpha R}$ are absent in the SM by construction, and thereby SM neutrinos are massless. However, the observation of neutrino oscillations in astrophysical and laboratory experiments implies that neutrinos have a mass~\cite{GonzalezGarcia:2007ib}. Even though the SM structure of the neutrino sector must be extended to accommodate the mass term, the neutrinos as indirect dark matter signals originate in charged and neutral current interactions of the left-handed fields $\nu_{\alpha L}$. As such, the effective operators which (via dark matter decay)
might lead to  high- and ultrahigh-energy neutrino lines in the energy spectrum need to involve $L_\alpha$. As an illustration, in Table~\ref{tab:1} we list hypothetical dark matter candidates, defined by standard model $SU(2)_L$ and $U(1)_Y$ quantum numbers, and the decay operators that would produce a neutrino line signal~\cite{Feldstein:2013kka}. All in all, the effective operators given in Table~\ref{tab:1} imply that neutrinos as indirect dark matter signals will always be accompanied by electromagnetic signals; e.g. secondary electrons will transform into photons scattering off the cosmic microwave background via the inverse Compton process. Neutrinos can be also produced through the decay of $\pi^\pm$ if the dark matter particle couples to $q\bar q$, but a photon counterpart will emerge from the associated $\pi^0 \to \gamma \gamma$ decay. Generally speaking, the assumption of a dominant neutrino channel carries with it a violation of the $SU(2)_L$ invariance, so as to allow a suppression of the $\ell_{\alpha L}$ coupling. However, exceptions could be manufactured, e.g., by allowing dark matter to decay into the sterile neutrino states $\nu_s$ (responsible for the generation of neutrino masses and lepton flavor mixing)  which can later mutate into active neutrinos $\nu_s \leftrightharpoons \nu_{\alpha L}$~\citep{Berezhiani:2015fba, Anchordoqui:2021dls}. Alternatively, the neutrino channel could be maximized introducing new degrees of freedom which would act as portals into the hidden sector~\citep{Hiroshima:2017hmy, Blennow19}.
However, it is clear that all of these effective models are able to suppress the coupling to charged leptons at the expense of  additional parameters that regulate the mixing between the hidden and visible (SM) sectors. Moreover, even if at tree level the $\ell_{\alpha L}$ coupling can be suppressed, radiative corrections could in principle start an electroweak cascade with the production of charged leptons and gauge bosons~\cite{Berezinsky:2002hq}. The center of attention in  
our analysis will be indirect dark matter searches in the neutrino channel, but we should always keep in mind that, in general, the same region of the parameter space could be tested by gamma-ray and cosmic-ray detectors~\cite{Kachelriess:2018rty,Blanco:2018esa}. Strictly speaking, we concentrate on decays of spin-0 and spin-1 dark matter particles yielding a $\nu_{\alpha L} \bar \nu_{\alpha L}$ final state. To simplify notation, hereafter the active SM left-handed neutrinos of flavor $\alpha$ are denoted by $\nu$, and the scalar and vector dark matter particles by $\chi$. The interesting decay channel in our study is then $\chi \to \nu \bar \nu$.

\begin{table}[t]
\begin{center}
\begin{tabular}{ccccc}
\hline \hline
~~~~~~~Case~~~~~~~ & ~~~~~~~Spin~~~~~~~ & ~~~~~~~$SU(2)_L$~~~~~~~ & ~~~~~~~$U(1)_Y$~~~~~~~ & ~~~~~~~Decay Operator~~~~~~~\\
\hline
1. & 0 & 3 & 1 & $\bar{L}_{\alpha}^c \phi L_{\alpha}$ \\
2. & 1/2 & 0 & 0 & $\bar{L}_{\alpha} H^c \psi$  \\
3. & 1/2 & 3 & 0 & $\bar{L}_{\alpha} \psi^a \tau^a H^c$ \\
4. & 1/2 & 2 & $-1/2$ & $\bar{L}_{\alpha} F \psi$  \\
5. & 1/2 & 3 & $-1$ & $\bar{L}_{\alpha} \psi^a \tau^a H$ \\
6. & 1 & 0 & 0 & $\bar{L}_{\alpha} \slashed{V} L_{\alpha}$ \\
7. & 3/2 & 0 & 0 & $(\bar{L}_{\alpha} iD_\mu H^c) \gamma^\nu \gamma^\mu \psi_\nu$ \\
\hline
\hline
\end{tabular}
\caption{Dark matter candidates and the decay operators which could produce a neutrino line signal. Here,  $H$ denotes the SM  Higgs doublet, $\phi$, $\psi$, $V^\mu$ or $\psi^\mu$ denote the dark matter particle depending on whether it has spin 0, 1/2, 1, or 3/2, respectively. In case 4, $F$  denotes either $B_{\mu\nu} \sigma^{\mu\nu}$, $\tilde{B}_{\mu\nu} \sigma^{\mu\nu}$, $W^a_{\mu\nu} \tau^a \sigma^{\mu\nu}$ or $\tilde{W}^a_{\mu\nu} \tau^a \sigma^{\mu\nu}$, with standard textbook~\cite{Halzen:1984mc} notation in which $B_{\mu \nu}$ and $W^a_{\mu\nu}$ represent the field strength tensors of the SM $U(1)_Y$ and $SU(2)_L$ gauge fields. In cases 1,4, and 5, for which the dark matter particle carries hypercharge, a Dirac mass partner is required. In this table we have only listed operators of lowest dimension when considering all operators allowed for either member of the Dirac pair.} 
\label{tab:1}
\end{center}
\end{table}

The favored models of dark matter are those characterizing $\chi$ as a relic density of weakly interacting massive particles (WIMPs). A key assumption of this WIMP paradigm is that $\chi$ is a non-relativistic stable particle species whose abundance is set by their annihilations in the early universe~\cite{Lee:1977ua, Vysotsky:1977pe, Goldberg:1983nd, Steigman:1984ac}. For temperatures above the $\chi$ mass, $T \gg m_\chi$, the dark matter particles are thought to be in thermal equilibrium with the SM plasma. When the temperature drops below $m_\chi$, the abundance of $\chi$ begins to decrease exponentially and $\chi \chi \to f \bar f$ annihilation processes become inefficient, where (in the simplest models) $f$ denotes any particle of the SM. Eventually, for $T_{\rm fo} \sim m_\chi/20$, the dark matter comoving density freezes out. The WIMP relic abundance (that is the fraction of the critical density contributed by $\chi$ today) is inversely proportional to the thermally-averaged velocity-weighted cross section for WIMP annihilation (to all channels) calculated at freeze-out: $\Omega_\chi  \propto h^{-2}/\langle \sigma v \rangle_{\rm fo}$, where  $h$ is the dimensionless Hubble constant.  The proportionality constant, which is steered by the dynamics of thermal freeze-out, is found to be
$3 \times 10^{-27}~{\rm cm^3} \, {\rm s}^{-1}$~\cite{Gondolo:1990dk}. Now, for a pair of non-relativistic WIMPs annihilating with relative velocity $v$, partial wave unitarity dictates an upper bound: $\Omega_{\rm DM}  \geq 1.7 \times 10^{-6} \sqrt{m_\chi/T_{\rm fo}} \, (m_\chi/{\rm TeV})^2 \, h^{-2}$~\cite{Griest:1989wd}, which implies
$m_\chi \leq 110~{\rm TeV}$~\cite{Blum:2014dca}. Curiously, a stable particle species with a weak-scale mass and interaction strength is predicted to freeze-out of thermal equilibrium with a relic abundance that is comparable to the measured cosmological density of dark matter: $\Omega_{\rm DM} \simeq 0.1186(20) \, h^{-2}$~\cite{Zyla:2020zbs}.  This can be seen taking a weak cross section derived from dimensional analysis: $\sigma \sim g_\chi^4/(4\pi m_\chi)^2 \sim 10^{-8}~{\rm GeV}^{-2}$, with $m_\chi \sim 1/\sqrt{G_F}$, $g_\chi \sim 0.65$, and $v \sim c/3$ for $T_{\rm fo} \sim m_\chi/20$~\cite{Steigman:2012nb}. This remarkable coincidence is usually referred to as the ``WIMP miracle''.  Thus far, WIMPs have eluded detection through any of the methodologies listed above \cite{MarrodanUndagoitia16, Gaskins16, Buchmueller17, Penning18, Rappoccio19}, motivating the consideration of alternative models of DM. Some classes of DM models feature non-thermal production in the early universe \cite{Chang:1996vw,Kuzmin:1997jua,Chung98,Birkel:1998nx, Chung99, Faraggi:1999iu,Kuzmin:1999zk,Chung01,Coriano:2001mg,Kannike:2016jfs,DelleRose:2017vvz} and result in a DM mass of $\gg 110~{\rm TeV}$ that could produce ultra-high energy cosmic rays or neutrinos through DM interactions. Following~\cite{Beacom:2006tt, Yuksel07, Arguelles19}, we assume that the $\chi$ particles can still annihilate efficiently in the Galactic halo via $\chi \chi \to \nu \bar \nu$, but we will remain agnostic about the specifics of model building, and more generally how these dark matter particles would evade the unitarity bound.

In the high-energy range, gamma-ray and cosmic-ray observatories provide strong constraints on the dark matter annihilation cross section and the particle decay widths \citep[e.g.][]{Sarkar:2001se,Cafarella:2003cx, Murase:2012xs, Aloisio15, Cohen:2016uyg, Kalashev16, Alcantara19, Anchordoqui:2018qom, Anchordoqui:2021crl, Maity:2021umk}. Observatories sensitive to high- and ultrahigh-energy neutrinos, such as  IceCube \citep{Gaisser:2014foa}, ANTARES \citep{Antares:2011nsa}, the Pierre Auger Observatory (Auger) \citep{Aab:2019auo}, ANITA \citep{ANITAIV19} and in the future for instance IceCube-Gen2 \citep{IceCubeGen2}, KM3Net \citep{Adrian-Martinez:2016fdl}, POEMMA \citep{POEMMA_JCAP}, GRAND \citep{GRAND_WP}, RNO-G \citep{Aguilar21RNOG}, can provide unprecedented constraints for these channels in the high to ultra-high dark matter mass range $m_{\chi} \gtrsim 10^3\,{\rm GeV}$.  Several existing studies consider annihilation channels \citep[e.g.][]{Arguelles19} or decay channels \citep[e.g.][]{Gondolo:1991rn, Esmaili12, Murase:2012xs, Rott:2014kfa, Cohen:2016uyg, IceCube18, Kachelriess18, Chianese21}, with various models for background neutrinos. For instance, the recent study by \cite{Chianese21} focuses on three decay channels, and on the IceCube, RNO-G and GRAND detectors, considering neutrino source and cosmogenic neutrino models as potential backgrounds. In this work, we calculate the sensitivities of POEMMA and GRAND, update the existing limits by Auger and ANITA by using the most up-to-date exposures, and we compare these sensitivities with existing limits from IceCube, with a particular emphasis on the uncertainties related to the dark matter spatial distribution in the Galactic halo. The layout of the paper is as follows. In Sec.~\ref{sec:distribution}, we describe the dark matter distribution and the neutrino intensity from decay or annihilation. In Sec.~\ref{sec:detectors}, we present key properties of the high and ultrahigh energy neutrino detectors considered. An observation strategy that can optimize the detection of neutrinos from dark matter decay or annihilation for POEMMA is presented in Sec.~\ref{sec:observation_strategy}. In Sec.~\ref{sec:annihilation} we present the constraints on the dark matter thermally averaged cross section and in Sec.~\ref{sec:decay} the constraints on the dark matter decay width. Their uncertainties are evaluated in Sec.~\ref{sec:uncertainties}. We discuss these prospective constraints and conclude in Sec.~\ref{sec:discussion}.

\section{Dark matter distribution and neutrino spectrum}\label{sec:distribution}

An accurate description of the dark matter distribution, in particular in the Galactic halo, is critical for direct and indirect searches. Its distribution is commonly assumed to be spherically symmetric and characterized by a specific radial profile, such as Navarro-Frenk-White (NFW) \citep{NFW97}, Burkert \citep{Burkert95}, or generalized NFW. The uncertainties concerning the shape of the profile as well as its normalization can be constrained by observations such as rotation curve measurements. In the following, in order to compare our estimates with estimates calculated by the IceCube Collaboration \cite{IceCube18}, we use the parameters given in \cite{Nesti13} for a Burkert profile
\begin{equation}
    \rho_\chi (r) = f_\chi \, \rho_H \left( 1+\frac{r}{R_H} \right)^{-1} \left[ 1+\left(\frac{r}{R_H}\right)^2 \right]^{-1}
\end{equation}
with a central dark matter density $\rho_H \simeq 4 \times 10^7 M_\odot \, {\rm kpc}^{-3}$ and a core radius $R_H \simeq 9 \, {\rm kpc}$. Here, $f_\chi$ is the fraction of dark matter that is superheavy. For comparison, and to account for the most recent estimates of the uncertainties related to the dark matter distribution \cite{Benito19, Benito20}, described in Secs.~\ref{sec:annihilation} and \ref{sec:decay}, we also consider a generalized NFW profile
\begin{equation}
    \rho_\chi (r) = f_\chi \, \rho_s \left( \frac{r}{R_s}
\right)^{-\gamma} \left( 1 + \frac{r}{R_s}
\right)^{-3+\gamma} \, , \end{equation}
where $\rho_s = \rho_0 \left( R_0/R_s \right)^{\gamma} \left( 1 + R_0/R_s \right)^{3-\gamma}$. The best fit parameters from \cite{Benito20} give a local density $\rho_0 = 0.6\,{\rm GeV\,cm}^{-3}$, a slope $\gamma = 0.4$ and a scale radius $R_s = 8 \times 10^1 \,{\rm kpc}$. The distance between the Sun and the Galactic Center is set to $R_0 = 8.178\,{\rm kpc}$ \cite{Gravity19}.

\begin{figure}[!ht]
    \centering
    \includegraphics[width=0.49\textwidth]{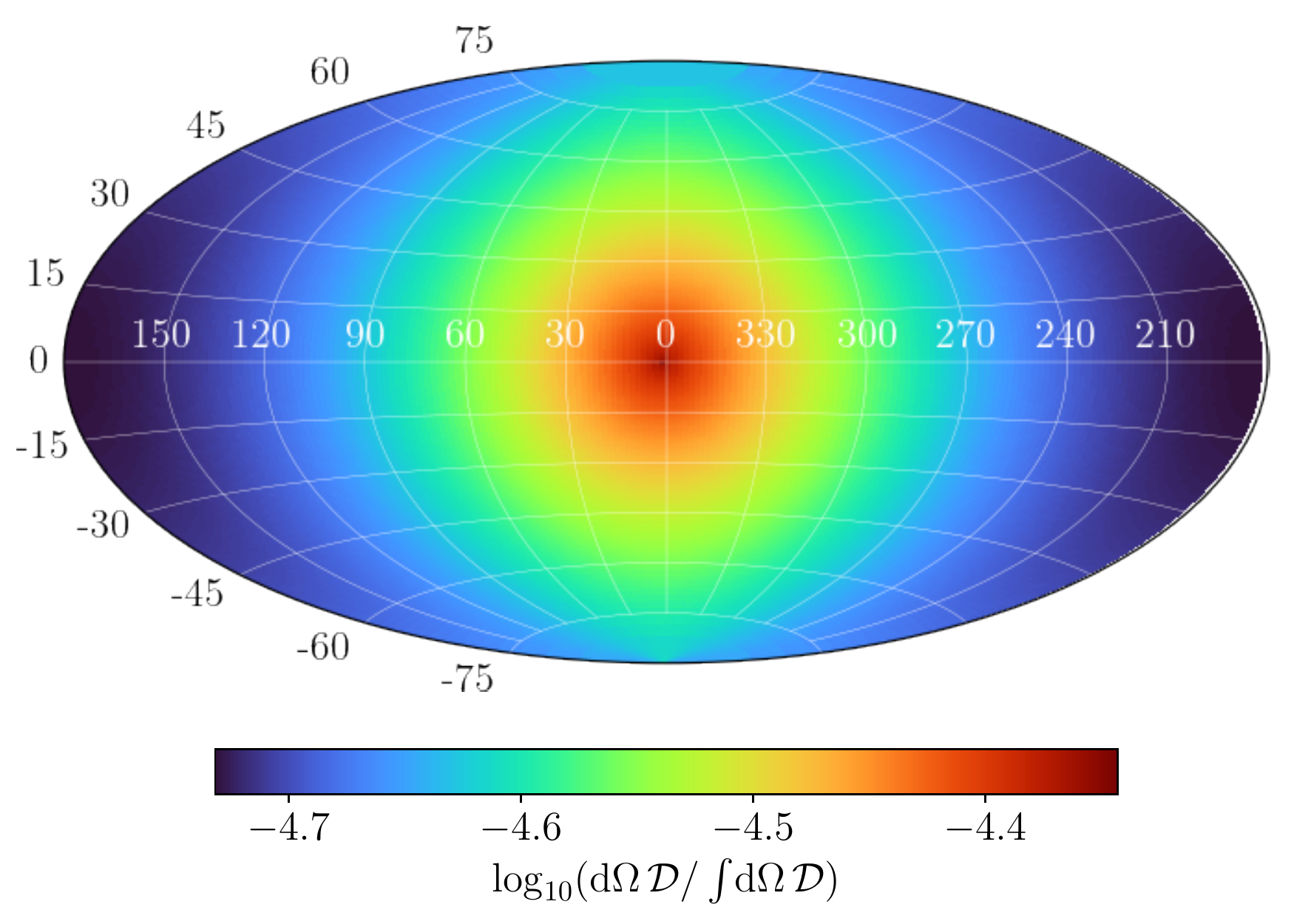}
    \includegraphics[width=0.49\textwidth]{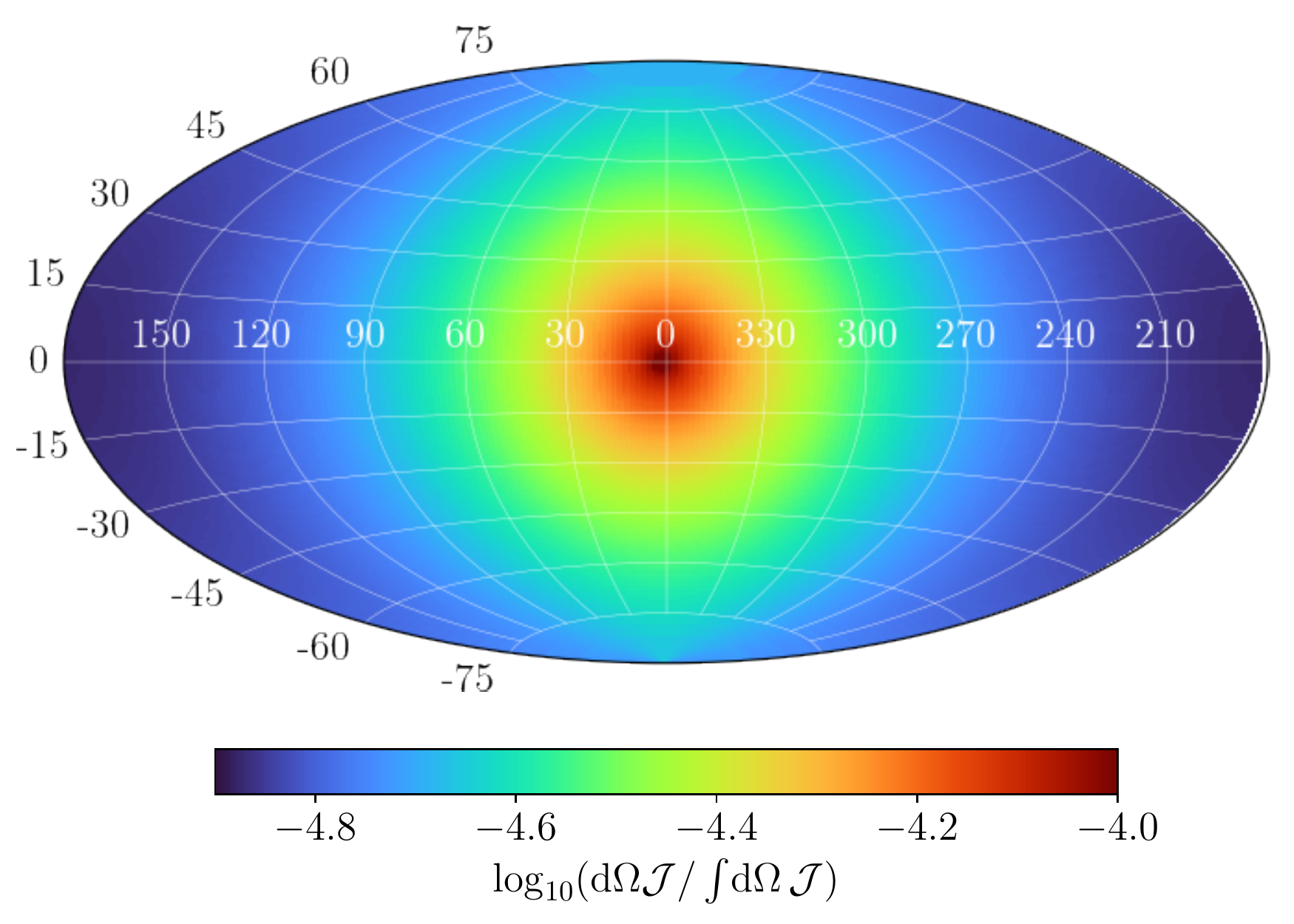}
    \caption{Differential $D$-factor for dark matter decay (left) and $J$-factor for annihilation (right) in different longitude and latitude bins for d$\Omega = 10^{-3}$ sr, considering the generalized Navarro-Frenk-White dark matter profile with best fit parameters from \citet{Benito20}. The $J$-factors are normalized by their integrals over the entire sky $\int {\rm d}\Omega \, {\cal D} = 6.6 \times 10^{23}\,{\rm GeV\,cm}^{-2}\,{\rm sr}$ and $\int {\rm d}\Omega \, {\cal J} = 2.3 \times 10^{23}\,{\rm GeV}^2\,{\rm cm}^{-5}\,{\rm sr}$, and are shown in logarithmic scales. }\label{fig:Jf_equatorial}
\end{figure}

Three dark matter astrophysical components contribute to the neutrino flux \citep[e.g.,][]{Beacom:2006tt,Yuksel07,Arguelles19}: the Milky Way halo, the extragalactic diffuse background and the extragalactic halos. In this work, we focus on the Milky Way halo component, as it provides stronger and less uncertain constraints than the ones provided by the Galactic center or extragalactic signals \citep{Yuksel07}. The average neutrino intensity in solid angle ${\rm d}\Omega$ from dark matter decay or annihilation \citep[e.g.,][]{Yuksel07,Leane20}
\begin{equation}\label{eq:intensity}
    \frac{{\rm d}\Phi}{{\rm d}\Omega {\rm d}E} \equiv \frac{\Gamma}{4 \pi m_\chi^a} \frac{{\rm d}N}{{\rm d}E} \, \int_{\rm l.o.s.} {\rm d}x \, \rho_\chi^a (x) \,,
\end{equation}
depends on the spectrum of decay or annihilation products ${\rm d}N/{\rm d}E$, the rate $\Gamma$, and the integral along the line of sight of the dark matter density, the so-called $D$-factor or $J$-factor
\begin{eqnarray}
    D(\Delta\Omega) &\equiv& \int_{\Delta \Omega} {\rm d}\Omega\ {\cal D}= 
    \int_{\Delta \Omega} {\rm d}\Omega \int_{\rm l.o.s.}{\rm d}x \, \rho_\chi(x)\\
    J(\Delta\Omega) &\equiv& \int_{\Delta \Omega} {\rm d}\Omega \ {\cal J}= 
    \int_{\Delta \Omega} {\rm d}\Omega \int_{\rm l.o.s.}{\rm d}x \, \rho^2_\chi(x)\ .
\end{eqnarray}
For decay, $\Gamma$ is the decay width and $a=1$. For annihilation, $\Gamma = \langle \sigma v \rangle /2$ with $\langle \sigma v \rangle$ the annihilation cross section, and $a=2$. Moreover, the factor $1/4\pi$ in Eq.~(\ref{eq:intensity}) accounts for isotropic emission. The line of sight distance $x$ and the galactocentric distance $r$ are related by $r = (R_0^2 - 2 x R_0 \cos\phi + x^2)^{0.5} $, with $\phi$ the angle between the line of sight and the Galactic center. The integral over $x$ is from 0 to the upper bound $x_{\rm max} = (R_{\rm halo}^2 - \sin^2 \phi R_0^2)^{0.5} + R_0 \cos\phi$, with $R_{\rm halo} = 30\,{\rm kpc}$. The differential $D$-factor and $J$-factor (${\cal D}$ and ${\cal J}$) are illustrated in figure~\ref{fig:Jf_equatorial}. For the decay and annihilation channels considered in this study, respectively $\chi \rightarrow \nu \bar{\nu}$ and $\chi \chi \rightarrow \nu \bar{\nu}$, the spectra of secondary decay or annihilation products peaks at $E_\nu = m_\chi/2$ and $E_\nu = m_\chi$, respectively. In the following, we use a delta-function approximation for these spectra (see Eqs.~\ref{eq:flux_nu_annihilation} and \ref{eq:flux_nu_decay}). We also assess the impact of the neutrino distribution \citep{Bauer20} in appendix~\ref{sec:neutrino_distribution}, for the case of decay.

\section{High and ultrahigh energy neutrino detectors}\label{sec:detectors}

A new generation of detectors, aiming at detecting ultrahigh energy particles and in particular very to ultra-high energy neutrinos (above $\sim 10^7\,{\rm GeV}$), is emerging. In this paper, we focus on the projects POEMMA (Probe of Extreme Multi-Messenger Astrophysics) and GRAND (Giant Radio Array for Neutrino Detection). Despite their common detection goal, these two future observatories involve different techniques and configurations. 

POEMMA will be comprised of two satellites flying in formation at $525\,$km altitude, equipped with Cherenkov and fluorescence detectors \citep{POEMMA_JCAP}. Cherenkov signals may come from extensive air showers from up-going $\tau$-lepton decays, the result of $\nu_\tau$ interactions in the Earth. A key feature is that the Earth acts as a neutrino converter. The probability for a $\tau$-lepton to emerge from the Earth and produce an up-going air shower depends on neutrino energy and its source location in the sky relative to the Earth, and the detectability depends on the satellites' positions \cite{Guepin19, Reno:2019jtr, Venters19}. Air fluorescence signals come from neutrino interactions in the atmosphere. Over several precession periods, POEMMA can access the full sky. In the Cherenkov observation mode, POEMMA can adopt specific observation strategies. For instance, the detectors can rapidly point toward a source in the case of an alert for a transient event. In its fluorescence detection mode, POEMMA will achieve a ground-breaking sensitivity to neutrinos in the range $\sim 10^{11}-10^{15}\,{\rm GeV}$. 

GRAND will be ground-based and composed of arrays of 10k to 200k radio antennas (referred to as GRAND10k and GRAND200k in the following) operating in the $50-200\,{\rm MHz}$ range in its final deployment \citep{GRAND_WP}. The targets for GRAND neutrino detection are also tau-leptons that decay to produce extensive air showers, coming from $\nu_\tau$ interactions in the Earth. A geomagnetic field effect yields
radio signals from the extensive air showers. The GRAND arrays can be deployed over immense areas and thus achieve a competitive diffuse sensitivity in the range $\sim 10^8-10^{11}\,{\rm GeV}$, together with a high duty-cycle in radio quiet areas. A single array of antennas will access a limited declination range. Full-sky coverage could be achieved by installing arrays at different locations around the globe. The latter configuration is still to be determined. 

Due to their prospective unprecedented neutrino sensitivity in the $>10^7$\,GeV energy range, these detectors are particularly well suited for constraining the neutrino production channels of superheavy dark matter. Several existing detectors sensitive to high to ultrahigh energy neutrinos already constrain indirectly the properties of superheavy dark matter. The properties of dark matter annihilating to neutrinos is constrained over a wide energy range considering various detectors in \cite{Arguelles19}. The properties of dark matter decaying into high-energy neutrinos is constrained for various experiments in \cite{Esmaili12, Kachelriess18}. These properties have also been constrained by the IceCube Collaboration for various decay channels \cite{IceCube18}. In this work we compare the sensitivities of POEMMA and GRAND with the constraints from IceCube, Auger and ANITA, which provide currently the most constraining limits from neutrino detection in the energy range considered.

\section{Observation strategies for POEMMA}\label{sec:observation_strategy}

\begin{figure}[t]
    \centering
    \includegraphics[width=0.49\textwidth]{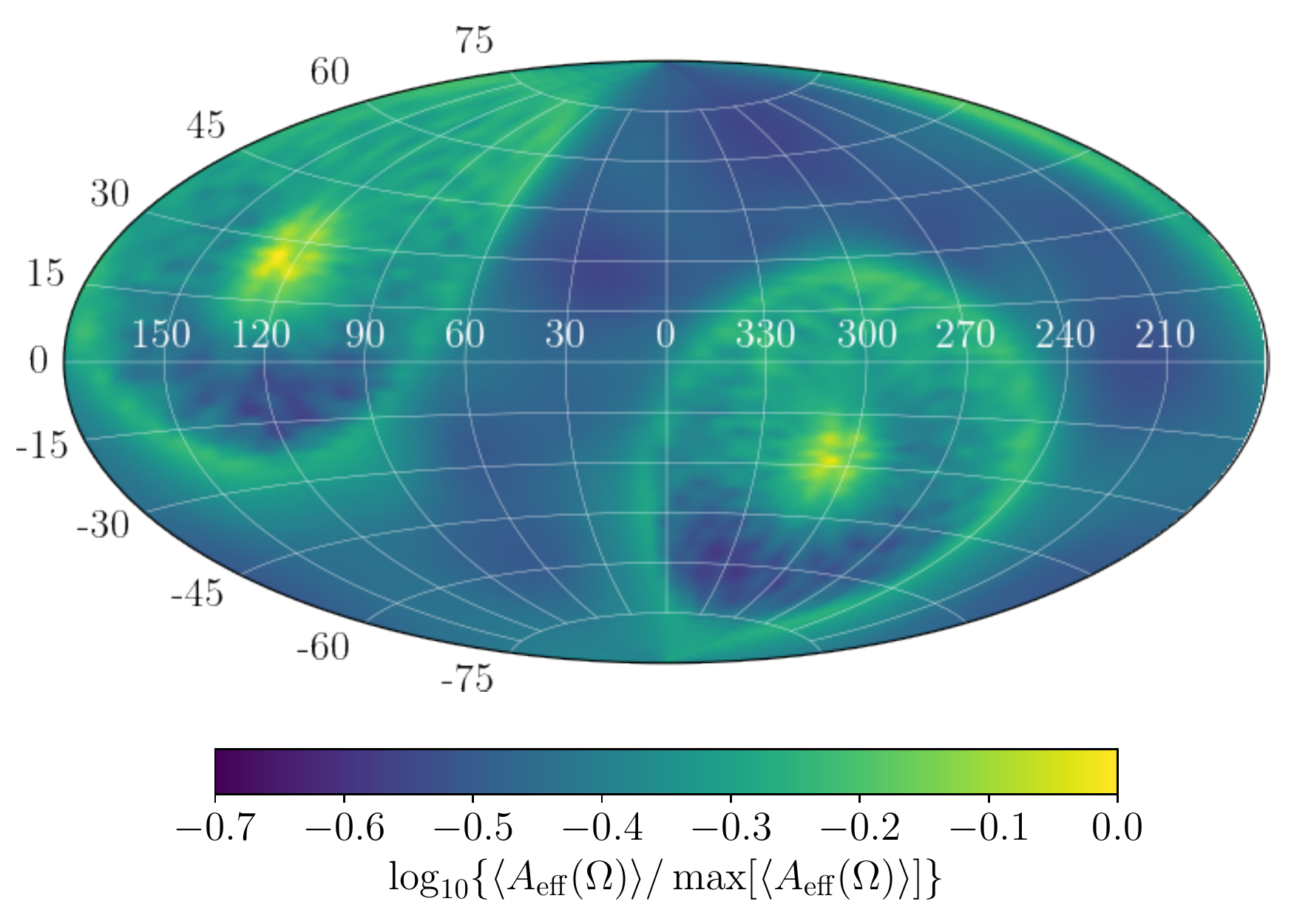}
    \includegraphics[width=0.49\textwidth]{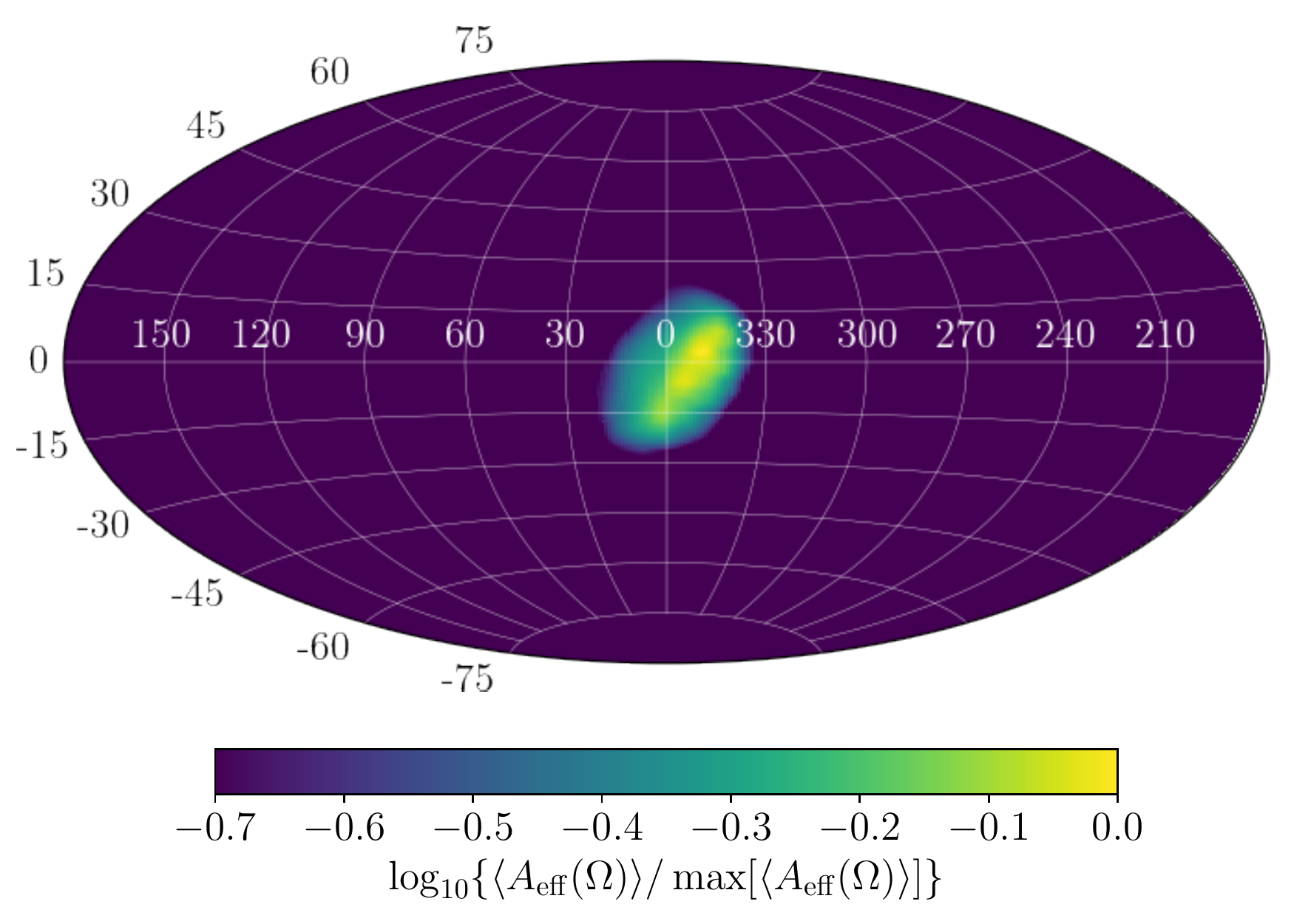}
    \caption{Normalized time averaged effective area in logarithmic scale, for the standard observation mode (left) and the Galactic center observation mode (right), for $E_\nu = 10^{8.5}\,{\rm GeV}$.}\label{fig:Exp_comp}
\end{figure}

Due to the slewing capability of its detectors, POEMMA can adopt various observation strategies in its Cherenkov observation mode. Full sky coverage can be achieved, and in the case of transient source follow-up, a specific observation strategy focussing on one region of the sky can be adopted \cite{Guepin19}. The dark matter density is enhanced in the Galactic center direction, which impacts the right ascension and declination dependencies or the differential $D$-factor and $J$-factor, as illustrated in Fig.~\ref{fig:Jf_equatorial}. The differential $J$-factor is the most impacted due to its dependency on $\rho_{\chi}^2$, against $\rho_{\chi}$ for the differential $D$-factor. Given these dependencies, an observation strategy optimized for indirect dark matter detection is important to develop. 

To determine the optimum observing strategy, we combine sky coverage calculations accounting for the detector field of view and orientation \citep{Guepin19} with calculations of the best achievable differential exposure for every direction of the sky \citep{Venters19}. In the sky coverage calculations, the detector has a field of view of $45\degree$ and covers a region
ranging from $7\degree$ below the limb to $2\degree$ above the limb. Also, we account for the illumination of the Sun and the Moon. For a total observation time $T_{\rm obs} \simeq 1\,{\rm yr} \; 15\,{\rm d} \; 4\,{\rm h}$, corresponding to $7$ precession periods of the satellite orbit around the north pole, we calculate the optimized effective observation time for every direction of the sky. To do so, we calculate the time-dependent detector orientation maximizing the effective area weighted by the ${\cal J}$-factor or ${\cal D}$-factor, for example, for DM annihilation, the quantity
\begin{equation}
    \int_{\Delta \Omega} {\rm d}\Omega \left. A_{\rm eff}(\Omega, E_\nu,t)
    \right|_{\rm max} \; \int_{\rm l.o.s.}{\rm d}x \, \rho^2_\chi(x) \, ,
\end{equation}
where $\left.A_{\rm eff}(\Omega,E_\nu,t)\right|_{\rm max}$ is the best achievable effective area for $\nu_\tau$ detection \cite{Venters19}, and $\Delta \Omega$ is the region of the sky determined by the instantaneous field of view of the detector \citep{Guepin19}. This procedure roughly corresponds to selecting the observable portion of the sky closest to the Galactic center.

The effective area depends on the area of the extensive air shower's Cherenkov cone subtended on the ground normal to the shower axis $A_{\rm Ch}(s)$, a quantity which
depends on the path length $s$ of the tau-lepton before its decay along a trajectory to the detector.  The effective area depends on the differential observation probability 
$dP_{\rm obs}$, according to 
\begin{equation}
    A_{\rm eff}(\Omega, E_\nu,t) = \int dP_{\rm obs}(\Omega,E_\nu,s,t)A_{\rm Ch}(s)\,,
\end{equation}
with the effective area averaged over $T_{\rm obs}$ designated by $\langle A_{\rm eff}(\Omega, E_\nu)\rangle$.
The differential probability to observe the shower depends on the probability of the tau-lepton to exit the Earth given an incident energy and angle of the tau neutrino, on the tau-lepton decay probability as a function of $s$ and on the detection probability given the shower energy, altitude and angle. Details can be found in \cite{Venters19}.
The effective observation time is normalized and used to weight the maximum effective area in every direction of the sky. In the following, this observation strategy is named Galactic center observation mode (GC), whereas the observation strategy leading to a full-sky coverage is referred to as standard observation mode (std). As illustrated in Fig.~\ref{fig:Exp_comp}, the two observation strategies lead to drastically different sky coverages.

\section{Dark matter annihilation to neutrinos}\label{sec:annihilation}

\begin{figure}[t]
    \centering
    \includegraphics[width=0.49\textwidth]{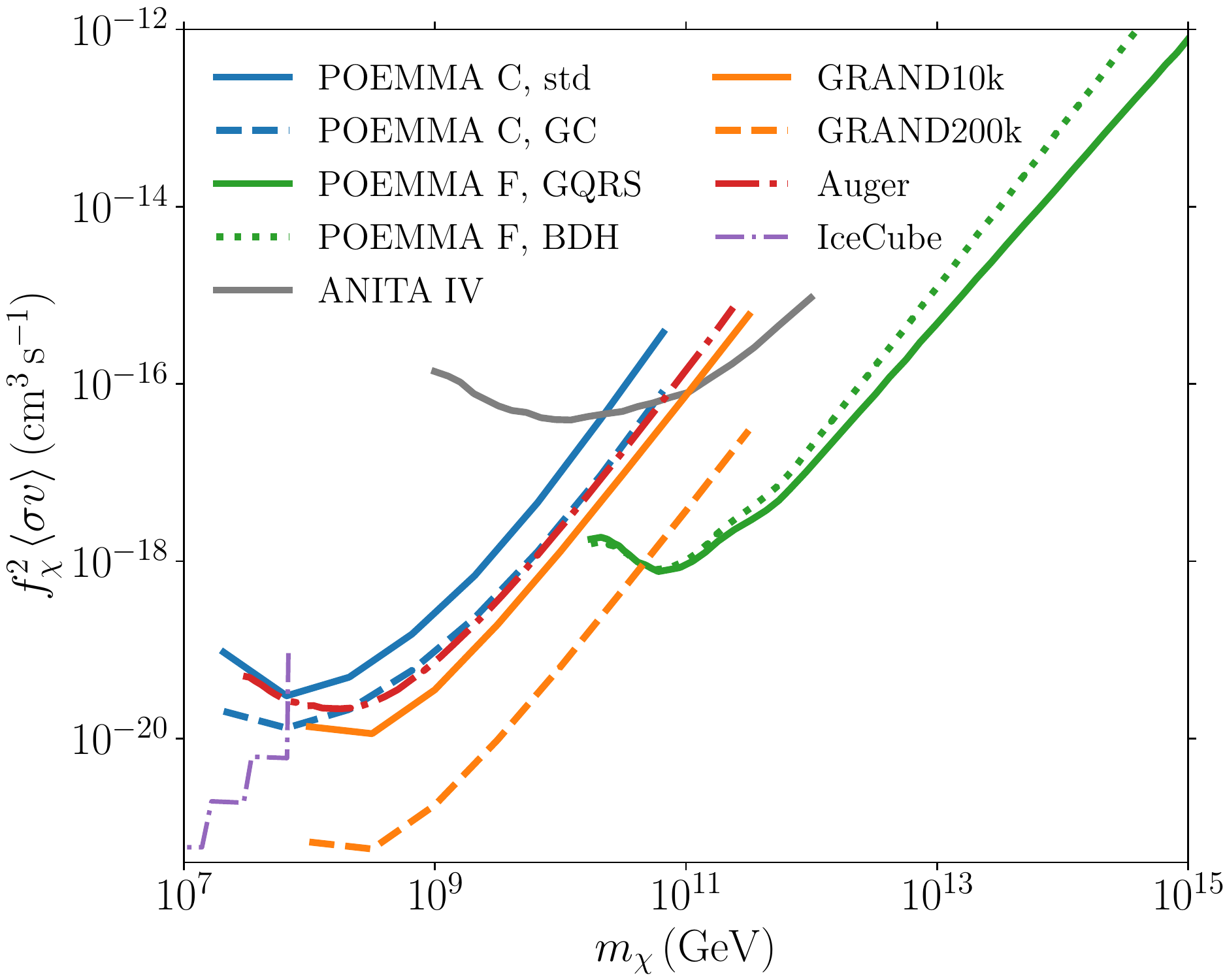}
    \caption{Sensitivities to dark matter thermally averaged annihilation cross section, $\nu\bar{\nu}$ channel (summed over neutrino flavors), multiplied by $f_\chi^2 = (\rho_\chi/\rho_{\rm DM})^2$. $5$-year sensitivities of POEMMA for the Cherenkov standard (std, solid blue) and Galactic center (GC, dashed blue), and the fluorescence (green) observation modes, GRAND10k (solid orange) and GRAND200k (dashed orange). Sensitivities of ANITA IV (grey), Auger (dot-dashed red), and IceCube \cite{Arguelles19} (dot-dashed purple).}\label{fig:CS_obs}
\end{figure}

In this section, we focus on dark matter annihilation to neutrinos, through the channel $\chi \chi \rightarrow \nu \bar{\nu}$, where $\chi$ is its own antiparticle with a cross section $\sigma\equiv \sum_i\sigma(\chi\chi\to \nu_\alpha \bar{\nu}_\alpha)$ for $\alpha =e,\ \mu,\ \tau$. We assume equal cross sections for each of the three neutrino flavors. For a given dark matter mass $m_\chi$, the three-flavor neutrino flux produced by dark matter annihilation in the Galactic halo is proportional to
a Dirac delta function at $E_\nu = m_\chi$
\begin{eqnarray}\label{eq:flux_nu_annihilation}
    \frac{{\rm d}\Phi_{\nu+\bar{\nu}}}{{\rm d}E_\nu}&=&\frac{1}{4\pi}\frac{\langle \sigma v \rangle}{2 m_\chi^2} [2\delta(m_\chi-E_\nu)] \int {\rm d} \Omega \int_{\rm l.o.s.} {\rm d}x \, \rho_\chi^2 (x) \, ,
\end{eqnarray}
where the factor $1/2$ accounts for the dark matter being its own antiparticle, and the factor of $2$ multiplying the Dirac delta function accounts for equal production of $\nu$ and $\bar{\nu}$. 
To account for the possible anisotropies of the sensitivity, or specific observation strategies and constrain the thermally averaged cross section, we combine POEMMA's
effective area \citep{Motloch:2013kva,Reno:2019jtr,Venters19} with the differential $J$-factor. The effective area is identical for neutrinos and antineutrinos for $E_\nu=m_\chi$ considered here. We refer to neutrinos and antineutrinos together as ``neutrinos'' in what follows. In terms of the time averaged effective area $\langle A_{\rm eff}(\Omega, E_\nu)\rangle$ and the observation time $T_{\rm obs}$, for a given annihilation cross section $\langle\sigma v\rangle (m_\chi)$, the number of detectable tau neutrinos at $E_\nu = m_\chi$ is given by
\begin{eqnarray}\label{eq:num_tau_annihilation}
    N_{\nu_\tau} (E_\nu) &=& \int {\rm d}E \, \frac{1}{4\pi}\frac{\langle \sigma v \rangle}{2 m_\chi^2} \frac{2\delta(m_\chi-E)}{\mathcal{N}_\nu} \int {\rm d} \Omega \int_{\rm l.o.s.} {\rm d}x \, \rho^2_\chi (x) \, \langle A_{\rm eff}(\Omega,E_\nu)\rangle\, T_{\rm obs}\, , \nonumber \\
    &=& \frac{1}{\mathcal{N}_\nu} \frac{1}{4\pi}\frac{\langle \sigma v \rangle}{E_\nu^2} \int {\rm d} \Omega \int_{\rm l.o.s.} {\rm d}x \, \rho^2_\chi (x) \, 
    \langle A_{\rm eff}(\Omega,E_\nu)\rangle\, T_{\rm obs}\,,
\end{eqnarray}
where $\mathcal{N}_{\nu} = 3$ is the number of neutrino flavors. POEMMA's Cherenkov signal sensitivities to the thermally averaged annihilation cross section multiplied by the square of the $\chi$-fraction of DM squared, $f_\chi^2 \langle \sigma v \rangle$, illustrated in Fig.~\ref{fig:CS_obs}, are given by setting $N_{\nu_\tau} = 2.44$, which corresponds to the $90\%$ C.L. limit with negligible background.

In the following we give additional detail about the calculation of the sensitivity for the different detectors considered. The number of detectable tau neutrinos is used to calculate the sensitivity for the Cherenkov observation mode of POEMMA, as noted above, and for GRAND10k and GRAND200k.  Total number of neutrinos,
$\sum_\alpha N_{\nu_\alpha}=2.44$, is used for the fluorescence observation mode of POEMMA, and for Auger and ANITA IV. 

For the Cherenkov observation mode of POEMMA, we use the averaged effective area
over a total observation time $T_{\rm obs}$ for the standard observation strategy, and the weighted effective area 
as described in section~\ref{sec:observation_strategy} for the Galactic center observation strategy. A detailed discussion of the prospective backgrounds for the Cherenkov observation mode of POEMMA can be found in \cite{Venters19}. In the cases of GRAND10k and GRAND200k, we use GRAND differential effective areas as a function of neutrino energy, for eight energy bins between $10^8\,{\rm GeV}$ and $10^{11.5}\,{\rm GeV}$. These differential effective areas are derived for an antenna array located at $43\degree$ latitude North (Olivier Martineau, private communication). The sensitivity calculated for GRAND200k (obtained by dividing the sensitivity of GRAND10k by $20$) is indicative, as the locations of the future twenty 10k antenna arrays are still to be determined.

In some of the cases considered, namely for the fluorescence observation mode of POEMMA, for Auger and for ANITA-IV, the differential exposure of the detector is not directly available in the literature. In these cases, we use the sensitivities of these detectors to compute the total exposure for one neutrino flavor $\mathcal{E} = 2.44 \, \mathcal{N}_\nu / [\ln(10) \, T_{\rm obs} \, 4\pi \, F_\nu ]$ where $\mathcal{N}_\nu = 3$ is the number of neutrino flavors, $T_{\rm obs}$ is the total observation time of the detector considered, and $F_\nu = E_\nu \, {\rm d} N_\nu / ({\rm d}E_\nu  \, {\rm d}A \, {\rm d} \Omega \, {\rm d}t)$ its sensitivity. The total exposure is then combined with the sky coverage of the detector to calculate the sensitivity to superheavy dark matter. For the fluorescence observation mode of POEMMA, its sensitivity \cite{Anchordoqui19, POEMMA_JCAP} using two different high energy neutrino cross sections, labeled GQRS \cite{GQRS98} and BDH \cite{BDH14}, is combined with a uniform differential exposure over the entire sky. The large instantaneous field of view of the fluorescence detector makes this assumption reasonable. In the case of Auger, the total exposure \citep{Aab15} multiplied by a factor $1.5$ to account for the increase of exposure with time, is combined with the average neutrino exposure per day \citep{Aab19} to account for the declination dependence of the exposure. In the case of ANITA-IV, its sensitivity \cite{ANITAIV19} is combined with the ANITA-III effective area as a function of declination \cite{ANITA20}.

The sensitivities computed can be compared with the limit from \cite{Arguelles19} for IceCube-HE (up to $10^8 \,{\rm GeV}$). As this limit is calculated using a generalized NFW dark matter profile, we simply scale it using the ratio $r_{J}$ between the full-sky $J$-factors $r_{J} = J_{\rm gNFW} / J_{\rm Burkert}$. A comparison between existing limits \citep{Arguelles19} for Auger is presented in appendix~\ref{sec:lim_auger}.

\section{Dark matter decay to neutrinos}\label{sec:decay}

\begin{figure}[t]
    \centering
    \includegraphics[width=0.99\textwidth]{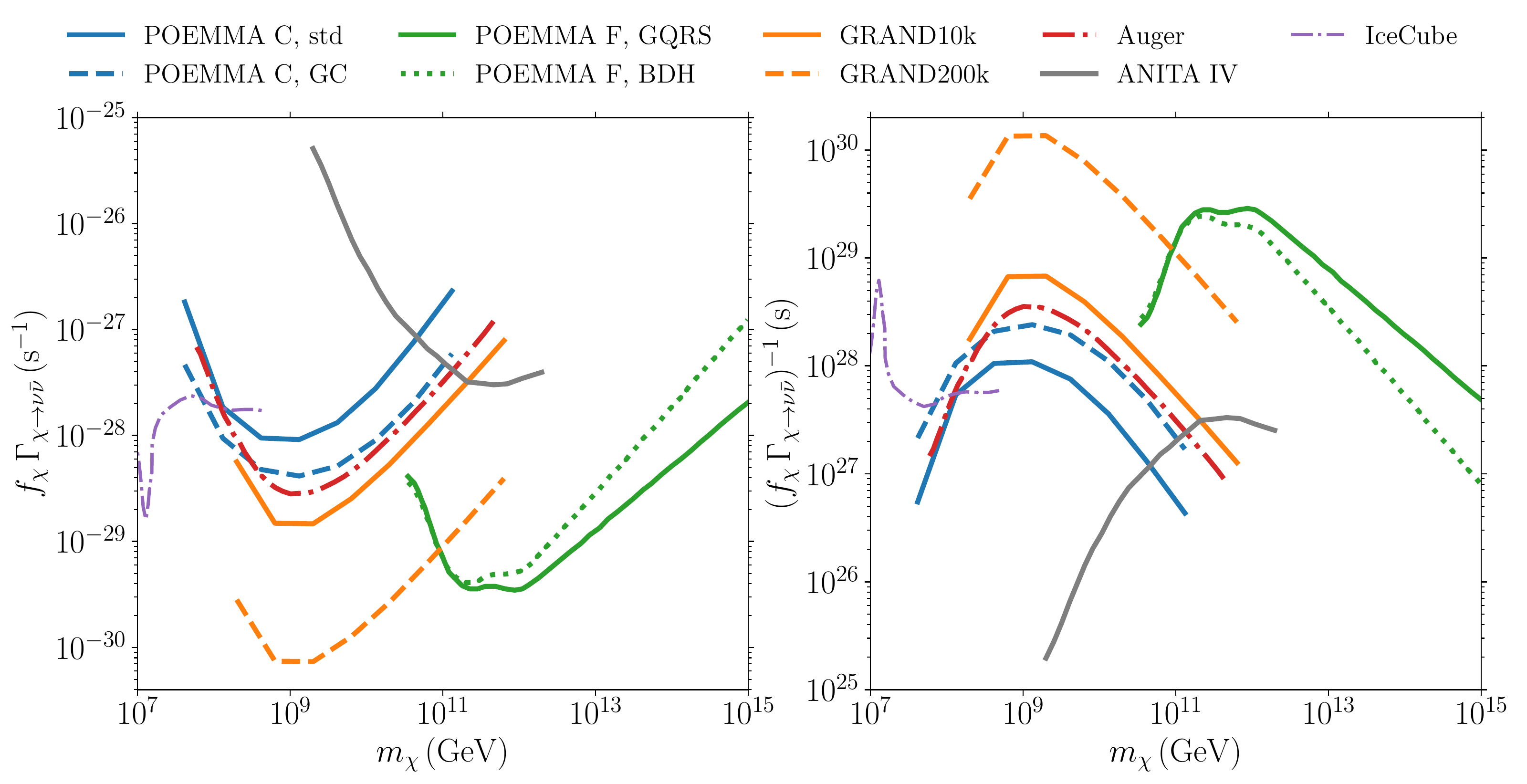}
    \caption{Sensitivities to dark matter decay width (left) and inverse of the decay width (right), $\nu\bar{\nu}$ channel. $5$-year sensitivities of POEMMA for the Cherenkov standard (std, solid blue) and Galactic center (GC, dashed blue), and the fluorescence (green) observation modes, GRAND10k (solid orange) and GRAND200k (dashed orange). Sensitivities of ANITA IV (grey), Auger (dot-dashed red), and the IceCube \cite{IceCube18} (dot-dashed purple). Allowed regions are below (above) the curves in the left (right) figure.}\label{fig:TAU_obs}
\end{figure}

Following the approach described in section~\ref{sec:annihilation}, we focus in this section on dark matter decay into neutrinos, through the channels $\chi \rightarrow \nu \bar{\nu}$. The three-flavor neutrino flux produced at $E_\nu = m_\chi/2$ by dark matter decay in the Galactic halo
\begin{eqnarray}\label{eq:flux_nu_decay}
    \frac{{\rm d}\Phi_{\nu+\bar{\nu}}}{{\rm d}E_\nu}&=&\frac{1}{4\pi}\frac{\Gamma_{\chi\rightarrow\nu\bar{\nu}}}{m_\chi} [2\delta(m_\chi/2-E_\nu)] \int {\rm d} \Omega \int_{\rm l.o.s.} {\rm d}x \, \rho_\chi (x) \, ,
\end{eqnarray}
depends on the dark matter decay width $\Gamma_{\chi\rightarrow\nu\bar{\nu}}$. As in Eq.~\ref{eq:flux_nu_annihilation}, the factor of $2$ multiplying the Dirac delta function accounts for equal production of $\nu$ and $\bar{\nu}$. The number of detectable tau neutrinos is given by 
\begin{eqnarray}
    N_{\nu_\tau} (E_\nu) &=& \int {\rm d}E \, \frac{1}{4\pi}\frac{\Gamma_{\chi \rightarrow \nu \bar{\nu}}}{m_\chi } \frac{2\delta(m_\chi/2-E)}{\mathcal{N}_{\nu}} \int {\rm d} \Omega \int_{\rm l.o.s.} {\rm d}x \, \rho_\chi (x) \, \langle A_{\rm eff}(\Omega,E_\nu)\rangle\, T_{\rm obs}\,, \nonumber \\
    &=& \frac{1}{\mathcal{N}_{\nu}} \frac{1}{4\pi}\frac{\Gamma_{\chi \rightarrow \nu \bar{\nu}}}{E_\nu} \int {\rm d} \Omega \int_{\rm l.o.s.} {\rm d}x \, \rho_\chi (x) \, \langle A_{\rm eff}(\Omega,E_\nu)\rangle\, T_{\rm obs}\,,
\end{eqnarray}
where $\mathcal{N}_{\nu} = 3$ is the number of neutrino flavors. The $90\%$ C.L. limit $N_{\nu_\tau} = 2.44$ gives the sensitivities to the dark matter decay width $f_\chi \Gamma_{\chi\rightarrow\nu\bar{\nu}}$ for POEMMA Cherenkov mode and GRAND, and $\sum_\alpha N_{\nu_\alpha}=2.44$ for POEMMA fluorescence mode, Auger and ANITA IV, which are illustrated in Fig.~\ref{fig:TAU_obs}. We overlay the limit calculated by the IceCube Collaboration \citep{IceCube18}, corrected to account for the difference of dark matter distribution used. As previously, for Auger, a comparison with existing limits \citep{Esmaili12, Kachelriess18} is presented in appendix~\ref{sec:lim_auger}.

\section{Dark matter distribution uncertainties}\label{sec:uncertainties}

The sensitivities presented in Figs.~\ref{fig:CS_obs} and \ref{fig:TAU_obs} are computed considering the Burkert dark matter distribution with parameters from \cite{Nesti13}, as mentioned in Sec.~\ref{sec:distribution}. However, due to the limited knowledge of the baryonic component of the Galaxy, the dark matter distribution is loosely constrained by rotation curve measurements, leading to significant uncertainties on dark matter properties \cite{Benito19, Benito20}.

In order to systematically evaluate the impact of these distribution on the sensitivities to SHDM annihilation and decay into neutrinos, we consider the general fit presented in \cite{Benito20}, that uses rotation curve measurements for the parameters $\rho_0$, $\gamma$, $R_s$ and $V_0$ (the circular velocity of the Sun) for a generalized NFW dark matter profile, with the latest estimates of the Galactic parameters \citep{Eilers19, Gravity19}. We use the likelihood profiles publicly available (\url{https://github.com/mariabenitocst/UncertaintiesDMinTheMW}), and we calculate the $1\sigma$ uncertainties on our sensitivities for four degrees of freedom, by considering parameters such that $\chi^2 - \chi^2_{\rm best \, fit} < 4.72$.

\begin{figure}[!ht]
    \centering
    \includegraphics[width=0.49\textwidth]{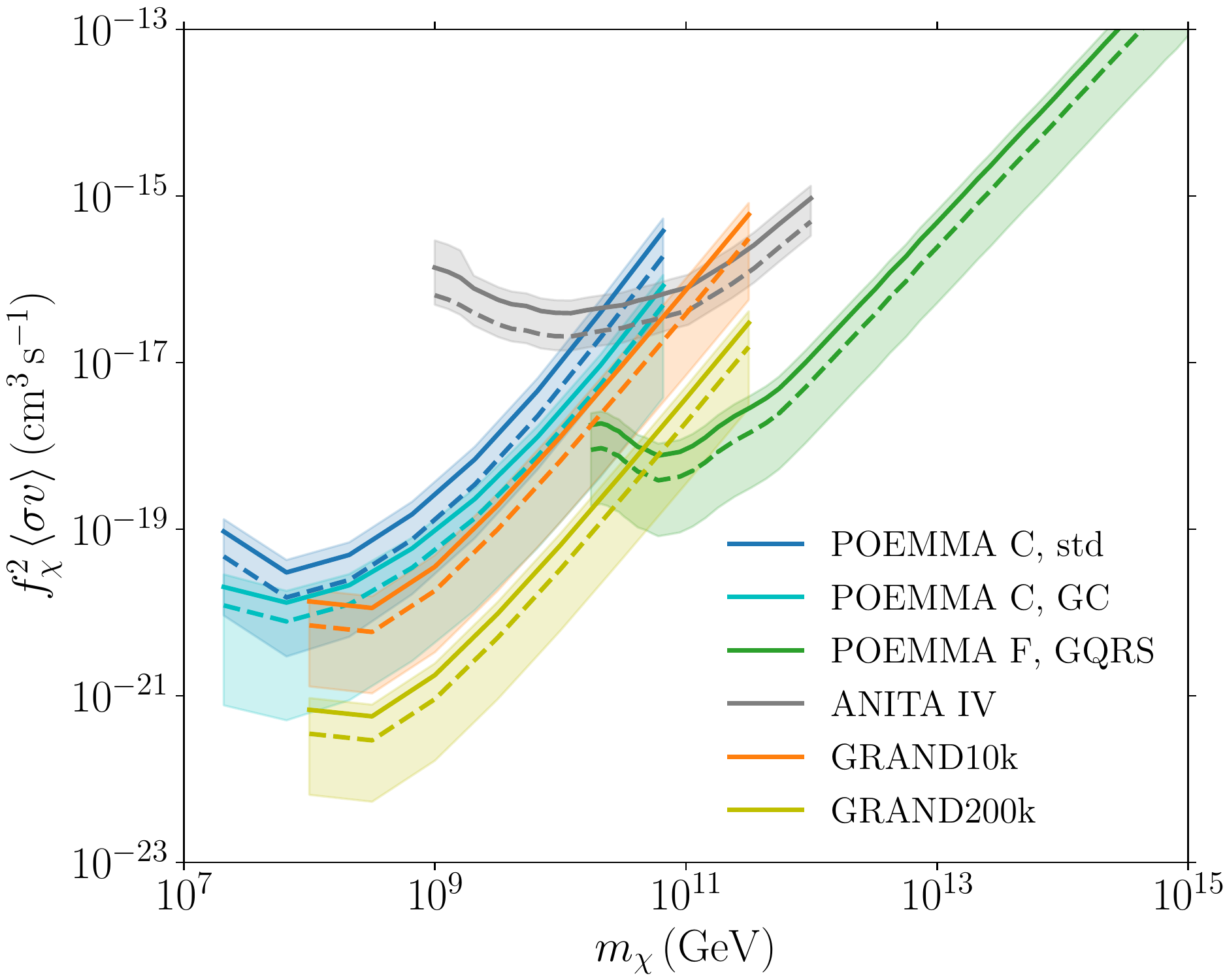}
    \includegraphics[width=0.49\textwidth]{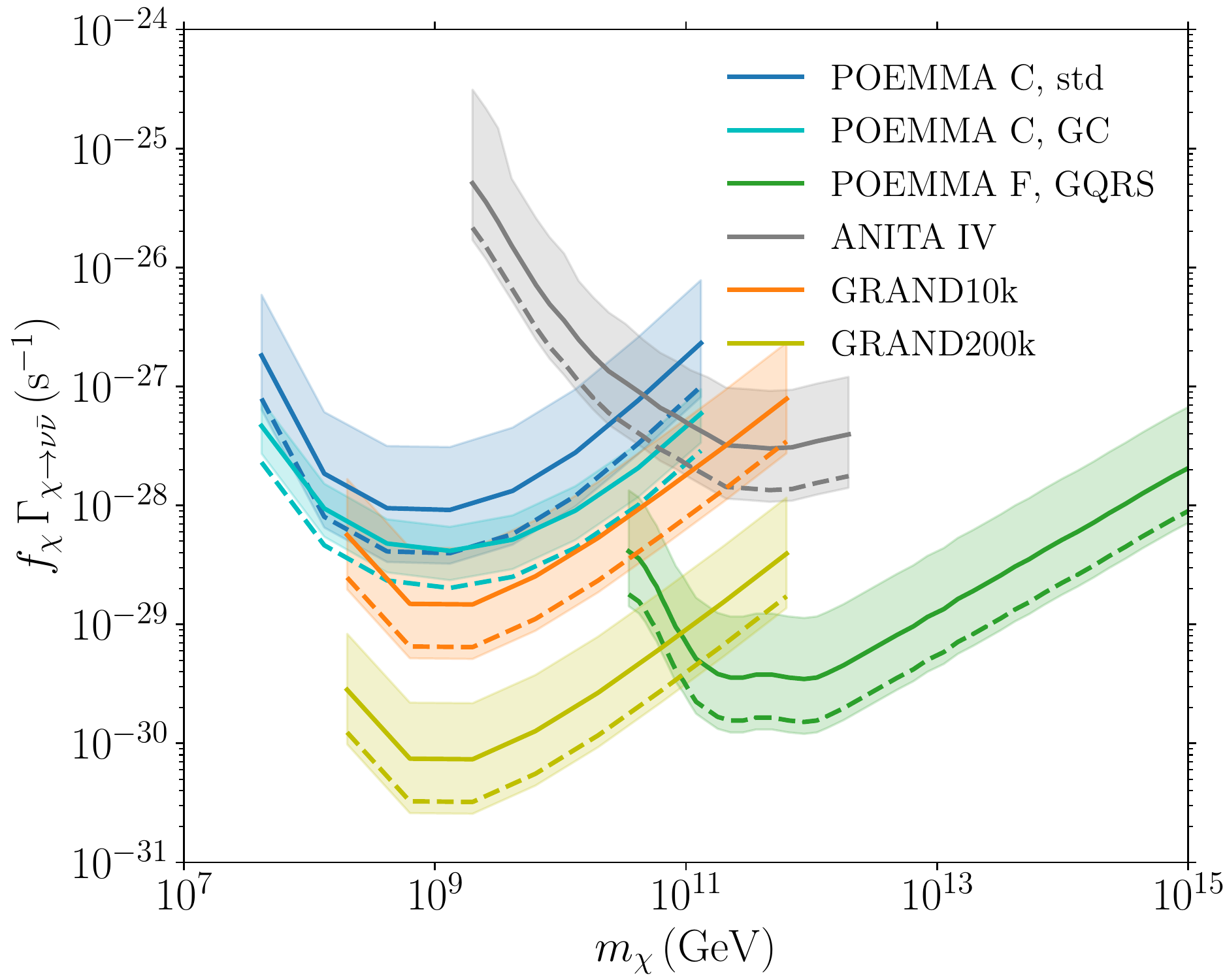}
    \caption{Uncertainties on the sensitivities to dark matter thermally averaged annihilation cross section (left) and on the sensitivities to dark matter decay width (right), for the $\nu\bar{\nu}$ channel. The bands show the $1\sigma$ uncertainties associated with the four parameters in the generalized NFW profile, the solid lines the sensitivities obtained with the Burkert dark matter distribution with parameters from \cite{Nesti13}, and the dashed lines the sensitivities obtained using the best fit parameters and the generalized NFW distribution from \cite{Benito20}.}\label{fig:CS_Gamma_uncertainties}
\end{figure}

The uncertainties to the sensitivities, in the case of annihilation and decay to neutrinos, are illustrated in Fig.~\ref{fig:CS_Gamma_uncertainties}. For both decay and annihilation, we obtain uncertainties of about $1-1.5$ orders of magnitude. As illustrated in Fig.~\ref{fig:Exp_comp}, the POEMMA Cherenkov Galactic center observation mode is only sensitive to the dark matter distribution in a restricted area around the Galactic center, and thus uncertainties are noticeably larger for the annihilation channel, due to the factor $\rho_\chi^2$ that intervenes in the calculation of the number of detectable neutrinos (see Eq.~\ref{eq:num_tau_annihilation}). Conversely, the ANITA experiment is mostly sensitive to a $\pm 20\degree$ declination band around ${\rm DEC} = 0\degree$ \citep{ANITA20}, and the uncertainties are smaller for the annihilation channel.

Estimates of the uncertainties due to the dark matter profile are available in the literature. In \cite{IceCube18}, the variation of the dark matter profiles can lead to uncertainties on the lifetime of the order of $\pm 10 \%$. These uncertainties are obtained for the Burkert model, by varying the parameters in the $1\sigma$ range \citep{Nesti13}, and for a comparison with the NFW model. In \cite{Arguelles19}, the likelihoods from \cite{Benito19} give uncertainties of approximately one order of magnitude.

\section{Motivation for future experiments}\label{sec:future_exp}

A space detector focussing on the Cherenkov detection of ultra-high energy neutrinos can be designed to observe a large portion of the limb, as a wide azimuth extent will increase the instantaneous sky coverage and thus the sensitivity of the detector. With a field of view of $45\degree$, POEMMA observes instantaneously approximately $1/12$ of the limb.

\begin{figure}[H]
    \centering
    \includegraphics[width=0.49\textwidth]{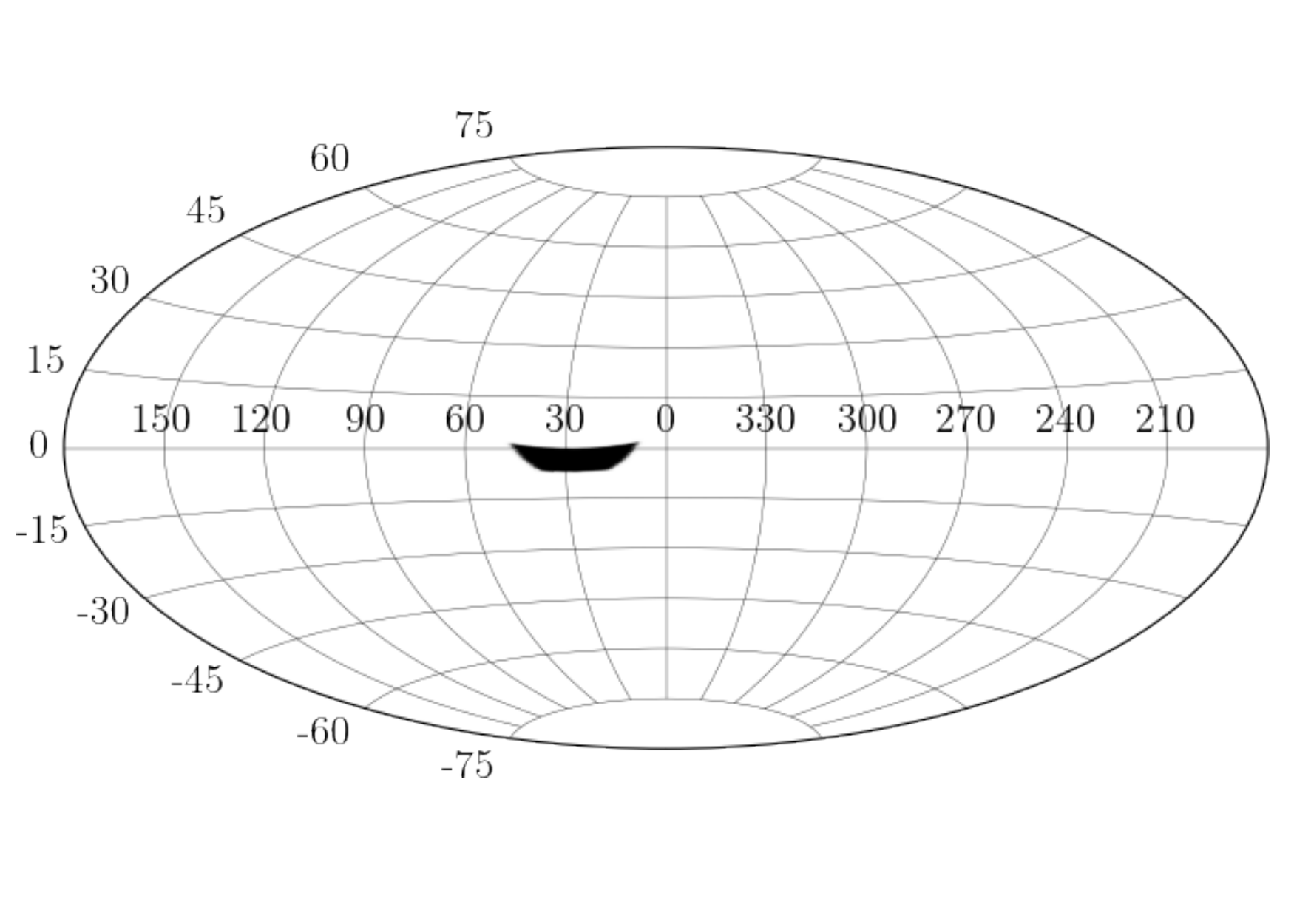}
    \includegraphics[width=0.49\textwidth]{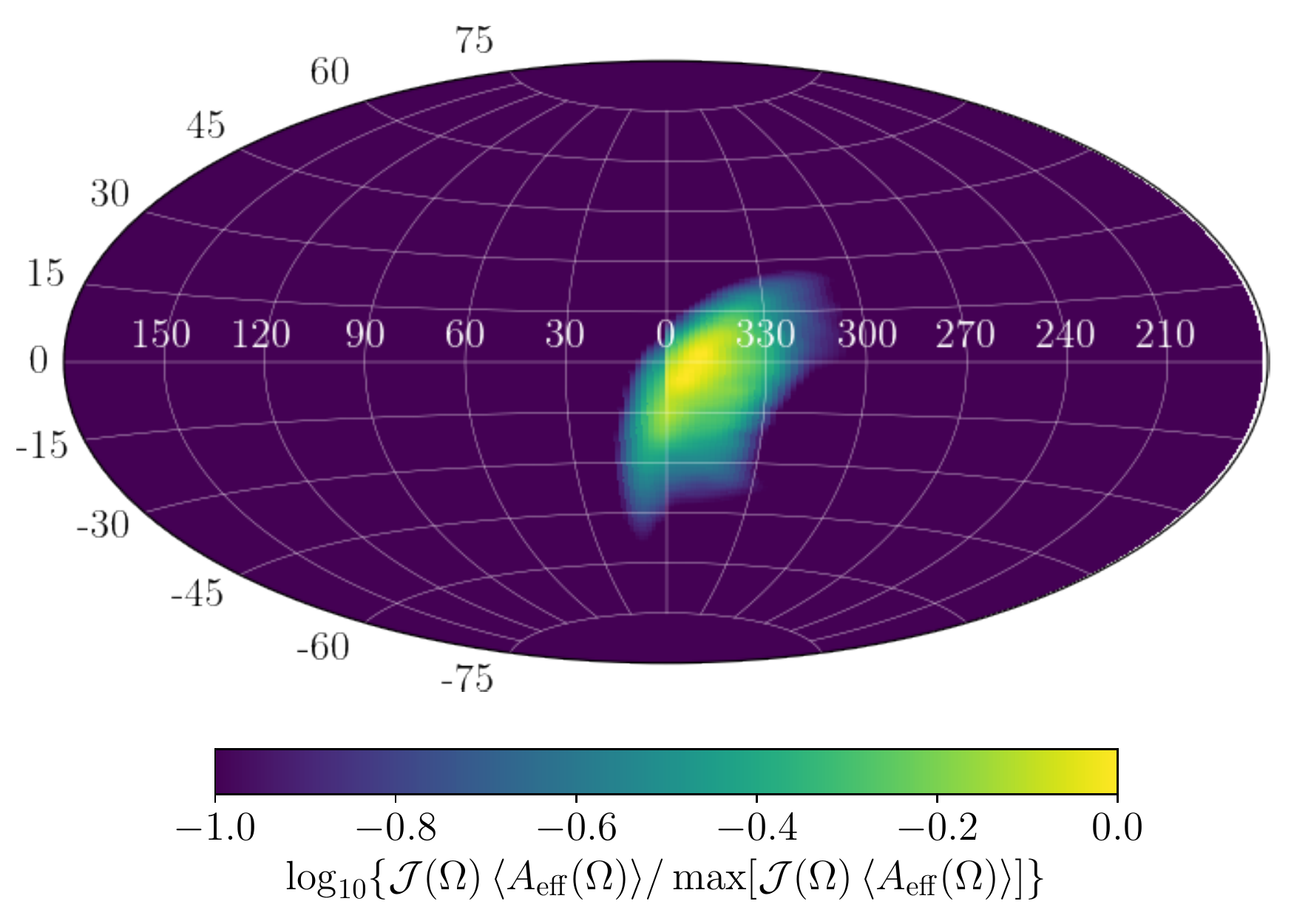}
    \includegraphics[width=0.49\textwidth]{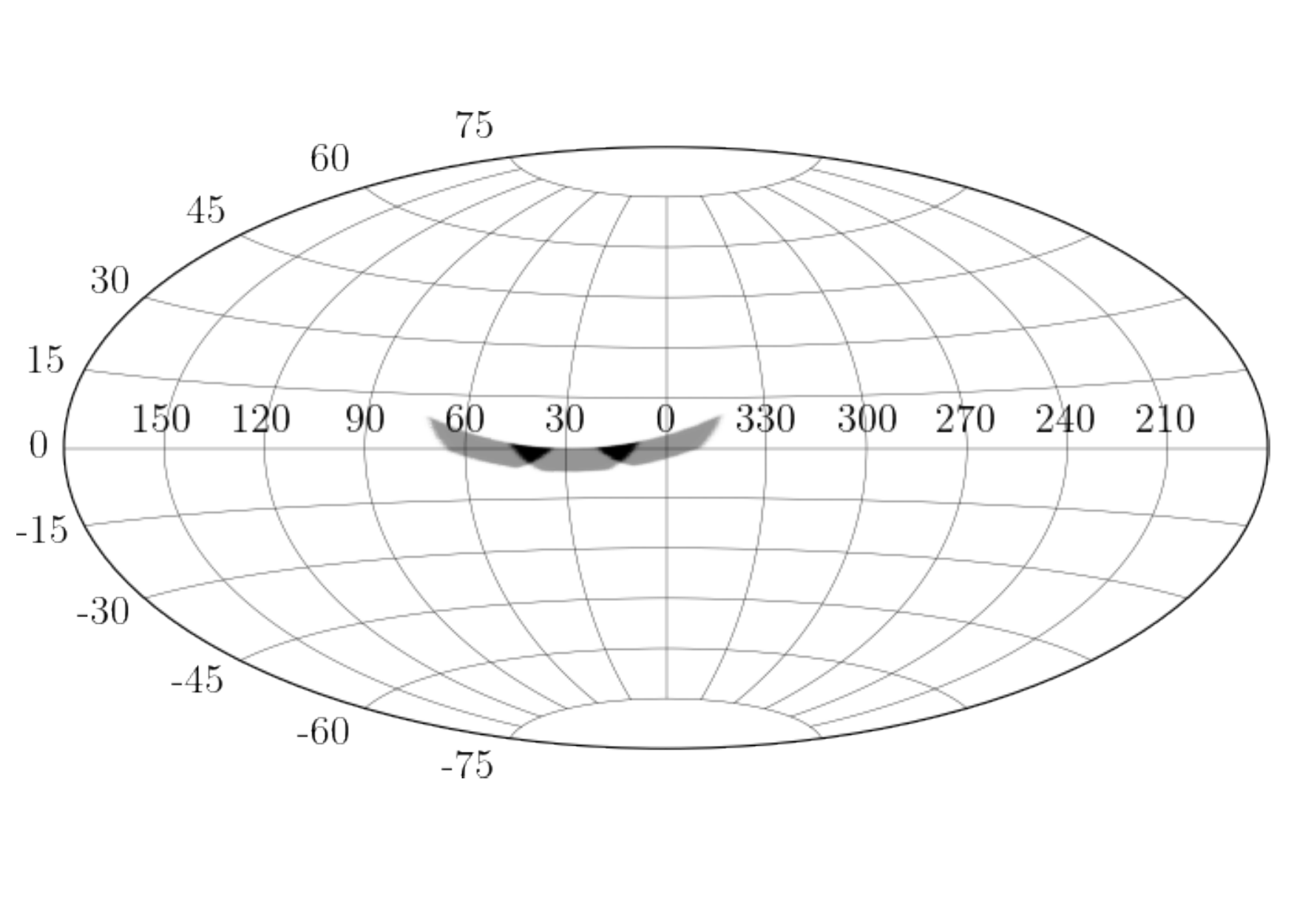}
    \includegraphics[width=0.49\textwidth]{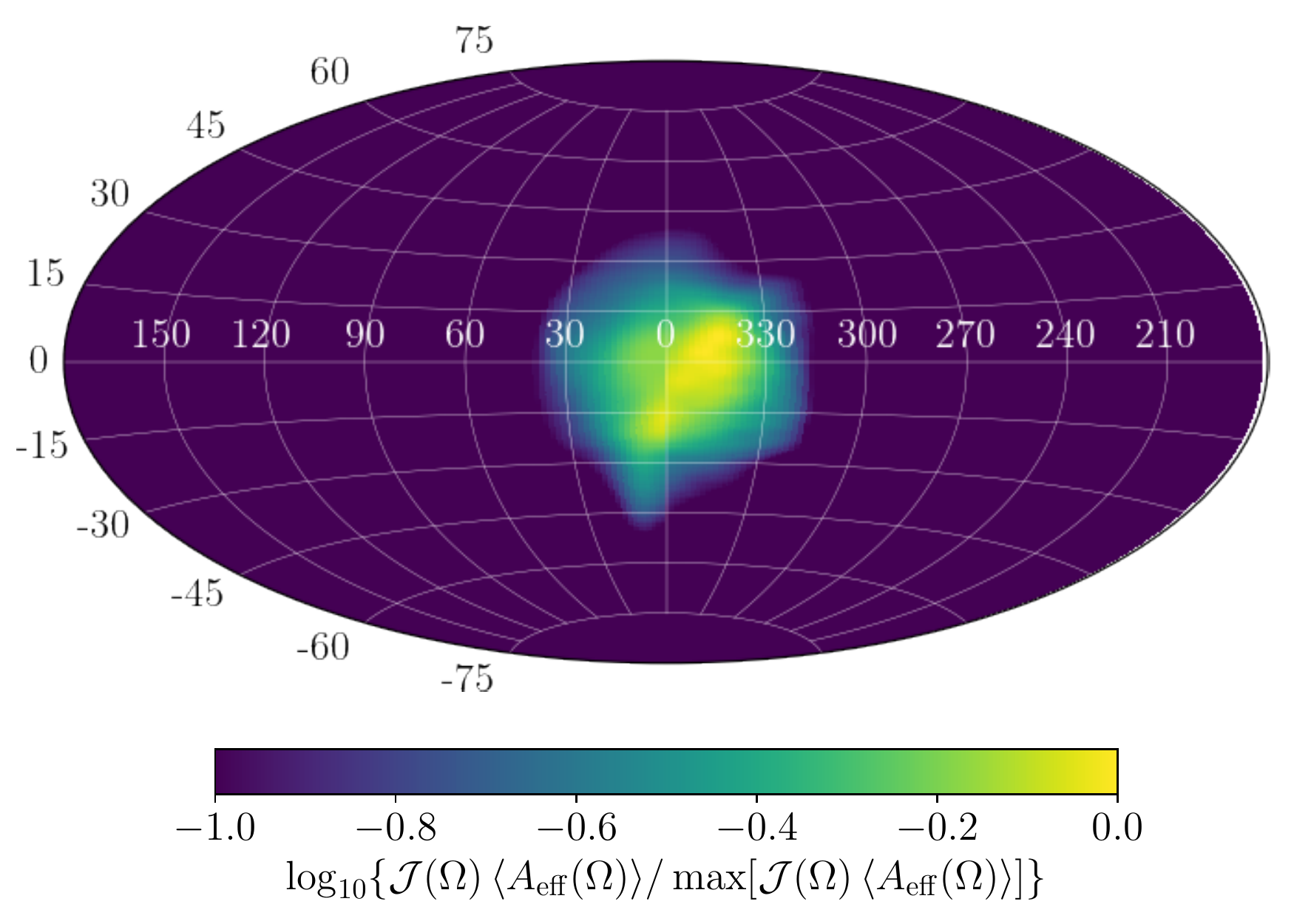}
    \includegraphics[width=0.49\textwidth]{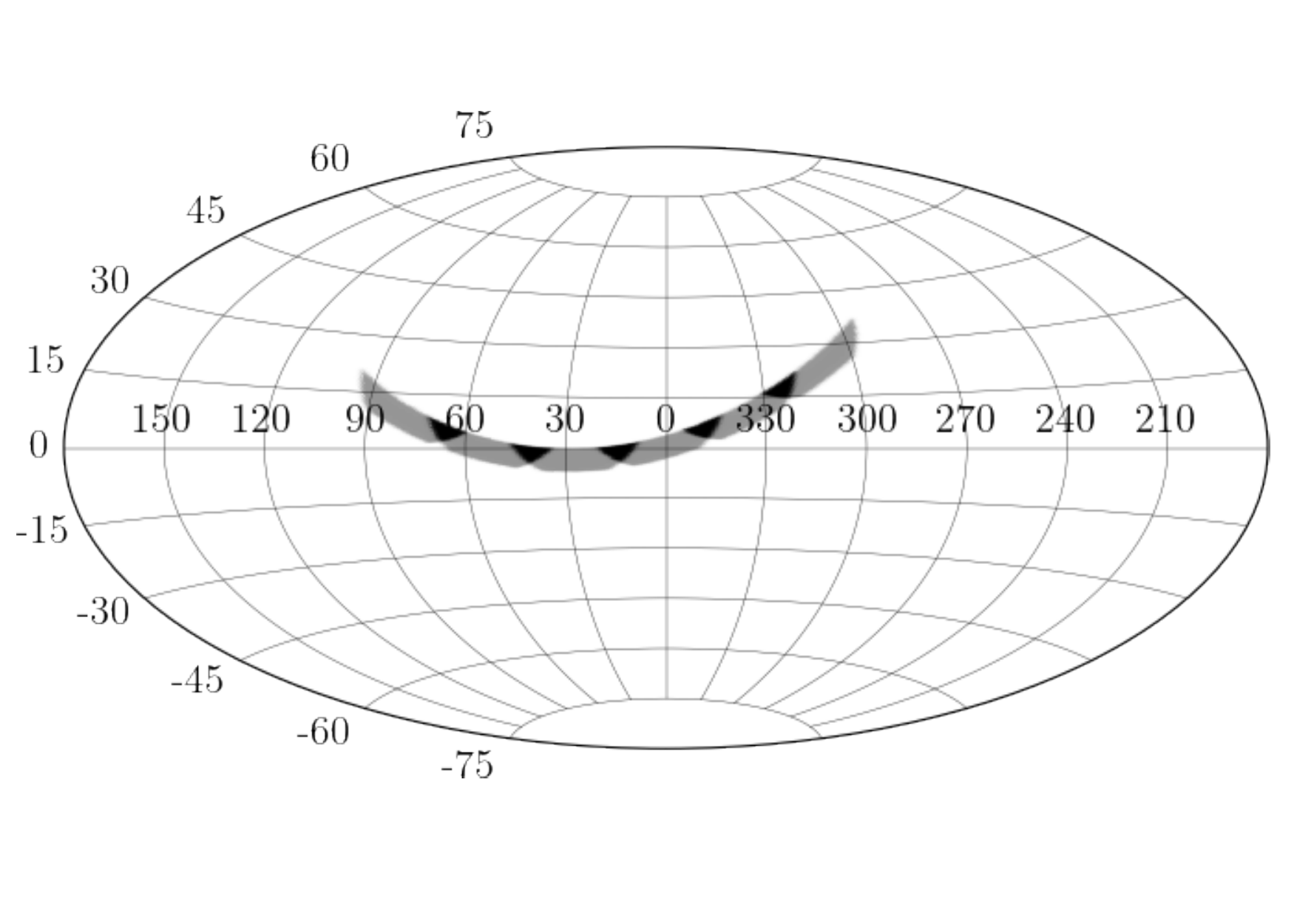}
    \includegraphics[width=0.49\textwidth]{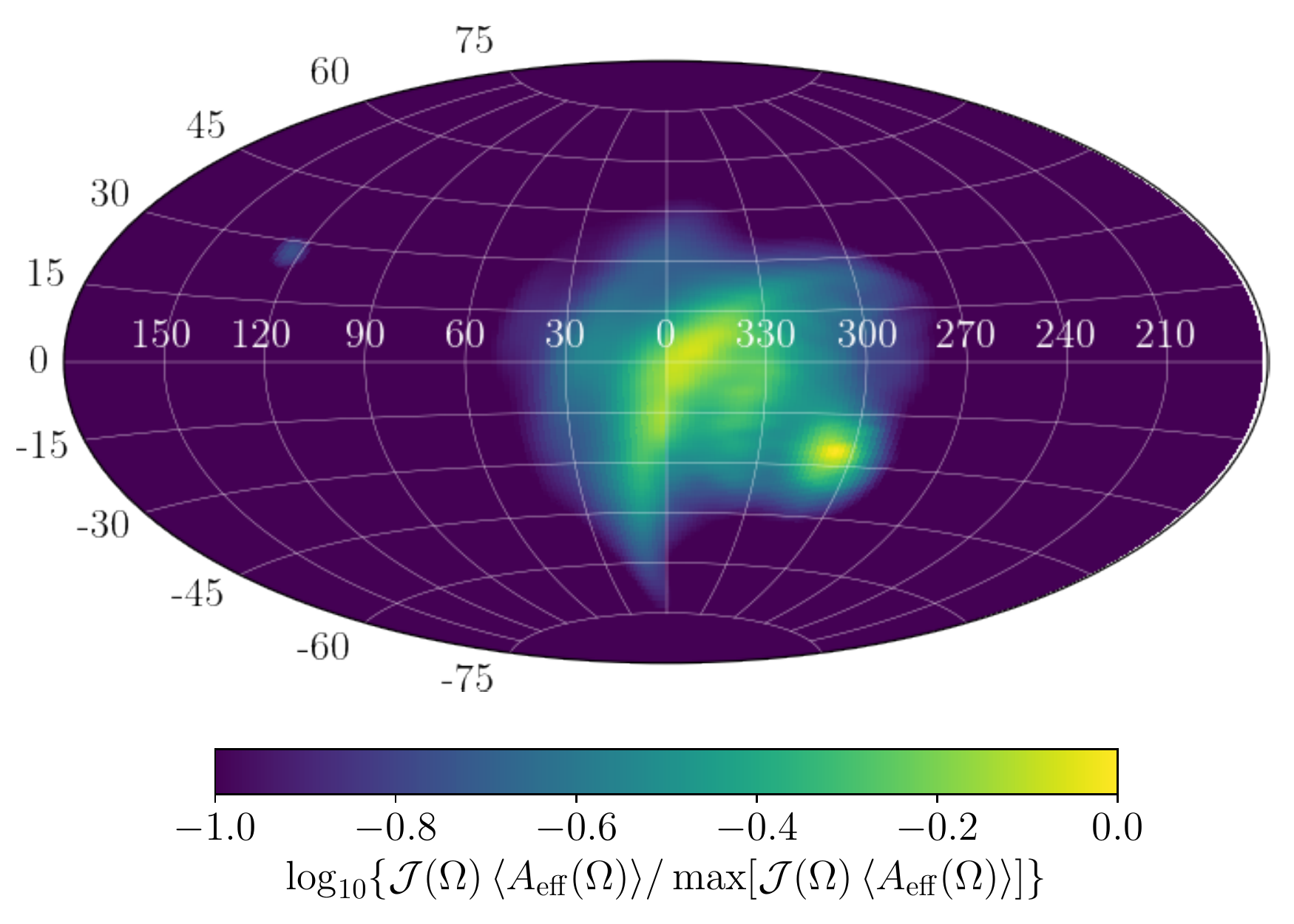}
    \caption{Instantaneous field of view for a given position along the satellite orbit (left) and effective areas weighted by the differential $J$-factor and normalized for $E_\nu  = 10^{8.5}\,{\rm GeV}$ (right), as a function of longitude and latitude, for configurations with one, three and six detectors (from top to bottom).}\label{fig:sky_cov_detectors}
\end{figure}

We evaluate the sensitivity gain that would be provided by a detector with a wider azimuth extent. Concretely, we consider several detectors pointing in different directions, all with a field of view $45\degree$, and covering a region ranging from $7\degree$ below the limb to $2\degree$ above. Three configurations are considered. The first is comprised of one detector, with an azimuth extent of $\sim 30\degree$, which corresponds to the POEMMA Cherenkov observation mode, the second is comprised of three detectors and has an azimuth extent of $\sim 90\degree$ and the third is comprised of six detectors with an azimuth extent of $\sim 180\degree$. The geometrical instantaneous fields of view of the three configurations are illustrated in Fig.~\ref{fig:sky_cov_detectors} for one satellite position along the orbit. The geometrical instantaneous field of view is given by the intersection between the region corresponding to the constraint on the viewing angle $\delta < 7\degree$ (or emergence angle $\theta_{\rm em} < 19.6\degree$), and the regions corresponding to the constraints on the field of view of the detectors ${\rm fov} = 45\degree$.

These three configurations are used to calculate the sky coverage of the instrument, using our optimization method accounting for the dark matter distribution. The central detector is pointed towards the direction maximizing the detection of dark matter, which is often the galactic center direction when accessible to observations. The effective areas weighted by the differential $J$-factor for the Burkert dark matter profile, namely $\mathcal{J}(\Omega) \, \langle A_{\rm eff}(\Omega,E_\nu)\rangle$, are illustrated in Fig.~\ref{fig:sky_cov_detectors} for $E_\nu  = 10^{8.5}\,{\rm GeV}$. The effective areas are time averaged over an observation time $T_{\rm obs} \simeq 1\,{\rm yr} \; 15\,{\rm d} \; 4\,{\rm h}$. Wider azimuth extents allow the detector to be sensitive to a larger portion of the sky, and the effect is more pronounced for the quantity $\mathcal{D}(\Omega) \, \langle A_{\rm eff}(\Omega,E_\nu)\rangle$ as $\mathcal{D}(\Omega)$ is less peaked towards the Galactic center direction than $\mathcal{J}(\Omega)$.

In order to evaluate the gain of the last two configurations when compared with the first configuration, for the case of annihilation we calculate and compare the quantities $\mathcal{I}_{J, n} = \int {\rm d}\Omega \, \mathcal{J}(\Omega) \, \langle A_{\rm eff}(\Omega,E_\nu)\rangle \, t_{\rm obs} (\Omega)$ for $E_\nu  = 10^{8.5}\,{\rm GeV}$, for the $1\sigma$ range presented in Sec.~\ref{sec:uncertainties}. In this formula, $n$ stands for the number of detectors, $\langle A_{\rm eff}(\Omega,E_\nu)\rangle$ is the best achievable effective area in all directions of the sky, weighted by the effective observation time for each direction $t_{\rm obs} (\Omega)$, which is computed using geometrical sky coverage calculations. For decay, we adopt the same procedure, calculating $\mathcal{I}_{D, n} = \int {\rm d}\Omega \, \mathcal{D}(\Omega) \, \langle A_{\rm eff}(\Omega,E_\nu)\rangle \, t_{\rm obs} (\Omega)$ at $E_\nu  = 10^{8.5}\,{\rm GeV}$.

For annihilation, the sensitivity gain is $\mathcal{I}_{J, 3}/\mathcal{I}_{J, 1} \sim 2$, and $\mathcal{I}_{J, 6}/\mathcal{I}_{J, 1} = 2-5$. For decay, the sensitivity gains also show a wide range with $\mathcal{I}_{D, 3}/\mathcal{I}_{D, 1} = 2-3$ and $\mathcal{I}_{D, 6}/\mathcal{I}_{D, 1} = 2-6$. These large ranges are due to the large uncertainties on dark matter halo properties. The best fit properties give gains of $\mathcal{I}_{J, 3}/\mathcal{I}_{J, 1} = 2$ and $\mathcal{I}_{J, 6}/\mathcal{I}_{J, 1} = 4$ for annihilation, and $\mathcal{I}_{D, 3}/\mathcal{I}_{D, 1} = 3$ and $\mathcal{I}_{D, 6}/\mathcal{I}_{D, 1} = 6$ for decay.

\section{Discussion and conclusion}\label{sec:discussion}

High to ultra-high energy neutrino detectors can provide unique constraints on the properties of superheavy dark matter annihilating or decaying to neutrinos. In this work, we have calculated the sensitivities and limits that high- to ultra-high energy neutrino observatories provide on dark matter thermally averaged annihilation cross section and dark matter decay width, for the channels $\chi \chi \rightarrow \nu \bar{\nu}$ and $\chi \rightarrow \nu \bar{\nu}$. We have focussed on calculating the sensitivities and limits of POEMMA, GRAND, ANITA-IV and Auger, compared with the current limits given by IceCube. The sensitivities of the detectors, or their differential exposures, as well as their sky coverages and the possibility of detecting several neutrino flavors, are key aspects for constraining the properties of superheavy dark matter.

The next stages of GRAND, GRAND10k and GRAND200k, to be deployed in the next decades, have the advantage of a very large exposure, of the detection from all azimuth angles and a full time operation due to the radio detection technique. Therefore they give the most constraining bounds in the energy range $\sim 10^8-10^{11}\,{\rm GeV}$. GRAND200k could improve the existing limits by two orders of magnitude. The locations of the antenna arrays of GRAND200k are still to be determined, which could influence its sensitivity to superheavy dark matter. The next phase of the experiment, GRANDProto300 \citep{Decoene19ICRC}, a preliminary network comprised of $300$ radio antennas, will determine the efficiency of autonomous radio detection and will possibly help identify unexpected sources of noise.

POEMMA has the advantage of full-sky coverage, due to its orbit around the Earth, and in the Cherenkov detection mode the pointing ability of the detector can allow optimizing the observation strategy for dark matter detection. A strategy focussing on the region of the sky observable and closest to the Galactic center improves the sensitivity of POEMMA to superheavy dark matter detection. This improvement is more significant for dark matter distributions peaked towards the Galactic center. At $10^9\,{\rm GeV}$, the sensitivity of POEMMA to superheavy dark matter decaying to neutrinos improves by a factor $\sim 2$ the constraint derived by the IceCube Collaboration \cite{IceCube18}. In the fluorescence observation mode, the three-flavor sensitivity and the full-sky coverage of POEMMA, lead to unprecedented sensitivity to superheavy dark matter properties above $10^{11}\,{\rm GeV}$, and improves by a factor of $\sim 80$ the sensitivity of ANITA-IV.

The uncertainties related to the dark matter distribution in the Galactic halo play a central role for indirect dark matter detection. In addition to calculating the sensitivities to superheavy dark matter annihilation and decay into neutrinos using the best-fit parameters of the dark matter distributions, we have evaluated the $1\sigma$ uncertainties on these sensitivities, using the tabulated uncertainties in the distribution of dark matter constrained from rotation curve measurements \citep{Benito20}. We have shown the importance of these uncertainties, that can be $1-1.5$ orders of magnitude, depending on the sky coverage of the detector considered.

An enhanced version of the POEMMA Cherenkov detector, for instance with a wider field of view, or comprised of several detectors pointing in different directions, could increase the sensitivity to superheavy dark matter properties. We consider the cases of three and six detectors with a field of view of $45\degree$. For the case of six POEMMA-like detectors, the best fit parameters of the generalized NFW distribution \citep{Benito20}, we find an enhancement in the sensitivity by a factor $4$ and $6$, respectively for annihilation and decay. Most of the detectors do not point towards the Galactic center, thus the enhancement is small for a very peaked dark matter distribution towards the Galactic center. Consequently, the uncertainties on the dark matter distribution strongly influence these estimates of the sensitivity gains.

In addition to GRAND and POEMMA, various projects of HE-UHE neutrino detectors are being developed, such as IceCube-Gen2 \citep{IceCubeGen2}, RNO-G, \citep{Aguilar21RNOG} Trinity \citep{Otte19}, and others \citep{Neronov19}, with a variety of detection techniques. These detectors will profitably contribute to superheavy dark matter searches.

\vspace{0.5cm}
{\bf Acknowledgements}

The authors thank Francis Halzen, Cosmin Deaconu and María Benito for useful discussions. C.G. is supported by the Neil Gehrels Prize Postdoctoral Fellowship. L.A.A. is supported by the U.S. National Science Foundation (NSF) Grant PHY-2112527. M.H.R. is supported in part by U.S. Department of Energy Grant DE-SC-0010113.

\appendix

\section{Distribution of secondary products}\label{sec:neutrino_distribution}

Electroweak showers can influence the distribution of secondary products for both the annihilation channel $\chi \chi \rightarrow \nu \bar{\nu}$ and decay channel $\chi \rightarrow \nu \bar{\nu}$ considered in this work. With SHDM, the $\nu$ or $\bar{\nu}$ can be produced with virtuality $\sim m_\chi$ ($m_\chi/2$) for annihilation (decay). Electroweak showers develop, degrading the initial neutrino energy, and through the showering, introduce additional neutrinos at lower energies. 
We consider recent calculations that include electroweak fragmentation function evolution, matching at the weak scale, and then further evolution with Pythia \citep{Bauer20}. We use the associated python packages including these effects,  available on github (\url{https://github.com/nickrodd/HDMSpectra}), to assess the impact of the distributions of secondary products on the sensitivities to SHDM. 

\begin{figure}[t]
    \centering
    \includegraphics[width=0.49\textwidth]{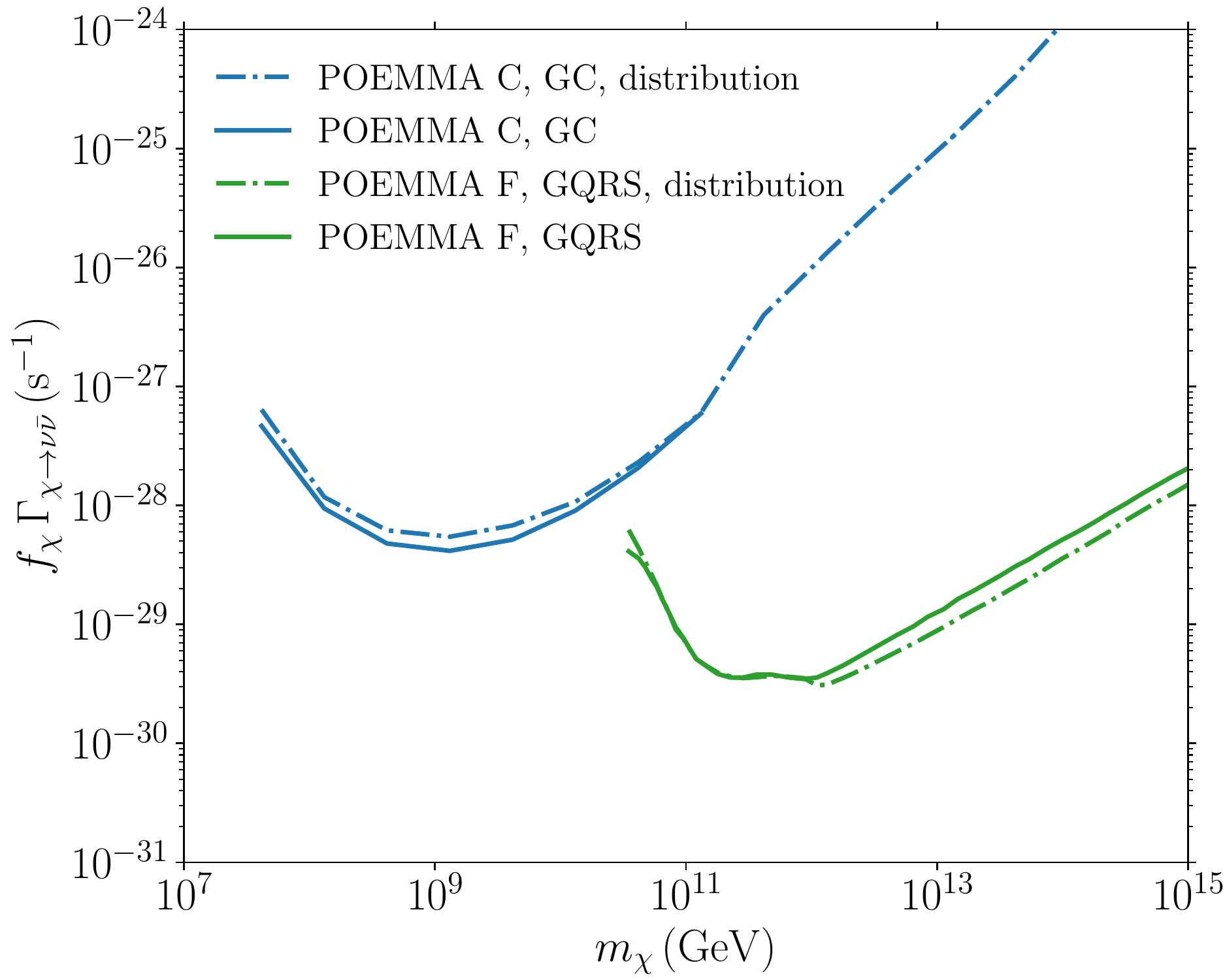}
    \caption{Effect of secondary neutrino distribution on the sensitivity to decay width for POEMMA Cherenkov (Galactic center observation mode) and POEMMA fluorescence (GQRS).}\label{fig:dist}
\end{figure}

Two examples for the decay channel, considering POEMMA Cherenkov Galactic center observation mode and POEMMA fluorescence observation mode with GQRS cross sections, are illustrated in Fig.~\ref{fig:dist}. For each mass $m_\chi$, with the inclusion of showering cascades, the number of neutrinos comes from the integral over the number of events in the POEMMA energy sensitivity range. The difference between the delta function approximation and the distribution including cascades is small when compared to uncertainties related to the dark matter distribution in the Galactic halo. A small enhancement appears at the highest energies due to the contribution of the low energy tail of the distribution of secondary products. The main difference appears for the Cherenkov observation mode, with a high-energy tail at $m_\chi > 1.4 \times 10^{11}\, {\rm GeV}$, which is produced by the low-energy tail of the neutrino distribution, thus without the contribution of the delta-function.

\section{Comparison with existing Auger constraints}\label{sec:lim_auger}
 
\begin{figure}[t]
    \centering
    \includegraphics[width=0.49\textwidth]{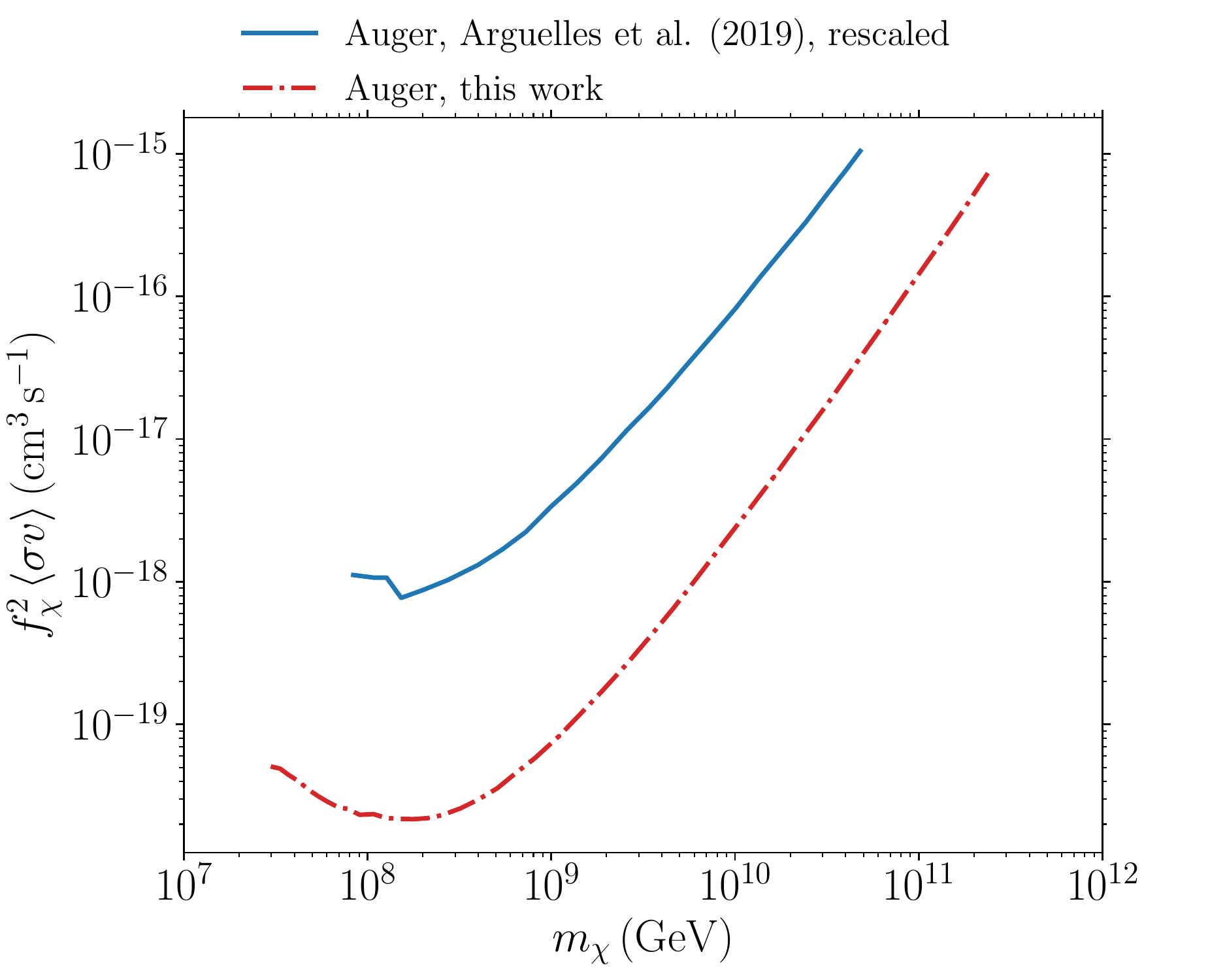}
    \includegraphics[width=0.49\textwidth]{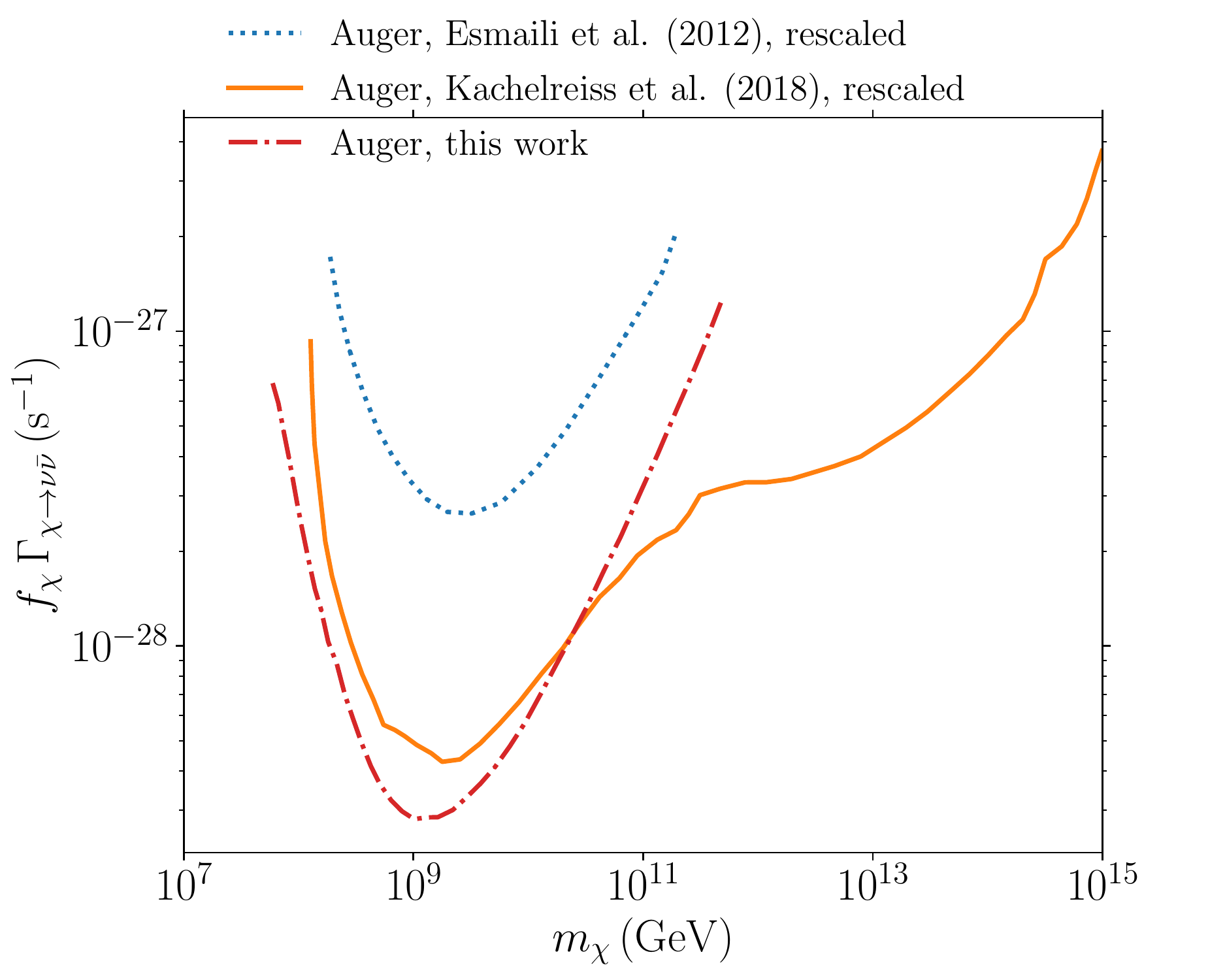}
    \caption{Comparison between Auger limits on the averaged annihilation cross section from \cite{Arguelles19} and this work (left) and Auger limits on the decay width from \cite{Esmaili12}, \cite{Kachelriess18} and this work (right).}\label{fig:lim_auger}
\end{figure}

Several constraints from the Auger experiment, on the SHDM annihilation cross section and decay width, have been computed in previous studies. In Fig.~\ref{fig:lim_auger} we compare our calculations, which use the updated Auger sensitivity to UHE neutrinos and the declination dependence of the day-average exposure \citep{Aab15, Aab19}, with estimates from \cite{Arguelles19} for the annihilation channel, and \cite{Esmaili12, Kachelriess18} for the decay channel. These estimates are rescaled to account for the different dark matter distribution profiles considered, and the increase of exposure with time.

Our limit for the annihilation channel differs by a factor $40$ from the rescaled constraint from \cite{Arguelles19}. The sky coverage of the detector and the related calculation of the differential $J$-factor differ in these two analyses.

For the decay channel, our limit differs from \cite{Esmaili12} (rescaled) by a factor of $10$, which may be related to different effective area and solid angle acceptance of the detector in both studies. Moreover, our estimate differs from \cite{Kachelriess18} (rescaled) by a factor of $2$ in the mass range $10^8-10^{11}\,{\rm GeV}$, the main difference between the two analyses being the use of the distribution of secondary neutrinos in \cite{Kachelriess18}, which contributes to the constraint in the higher mass range $10^{11}-10^{15}\,{\rm GeV}$.

\bibliographystyle{apsrev}
\bibliography{references}

\end{document}